\documentclass[a4paper,11pt]{article}

\usepackage[utf8]{inputenc}

\usepackage[backend=biber,style=numeric,maxnames=5,minnames=5]{biblatex}
\usepackage{xcolor}
\usepackage{graphicx}
\usepackage{amsmath,amsfonts,amssymb}
\usepackage{authblk}
\usepackage{booktabs}
\usepackage{comment}

\usepackage{subcaption}

\usepackage{hyperref}

\title{GEMINI: The First Underground Testbed for Seismic Isolation and Interplatform Control in Next-Generation Gravitational-Wave Detectors}
\author[1,2]{Tomislav Andric}
\author[1,2]{Jan Harms}
\author[1,2]{Ilaria Caravella}
\author[1,2]{Michele Angiolilli}
\author[2]{Daniele Cortis}
\author[2]{Nicola D’Ambrosio}
\author[2]{Massimiliano De Deo}
\author[2]{Marco D’Incecco}
\author[2]{Antonio Di Ludovico}
\author[3]{Oliver Gerberding}
\author[2]{Alessandro Lalli}
\author[5]{Brian Lantz}
\author[2]{Laura Leonzi}
\author[2,6]{Carla Macolino}
\author[7]{Richard Mittleman}
\author[8]{Conor Mow-Lowry}
\author[2]{Donato Orlandi}
\author[2]{Stefano Pirro}
\author[9, 10]{Marco Ricci}
\author[11]{Jamie Rollins}
\author[4]{Jim Warner}

\affil[1]{Gran Sasso Science Institute (GSSI), L'Aquila, Italy}
\affil[2]{INFN - Laboratori Nazionali del Gran Sasso (LNGS), Assergi, Italy}
\affil[3]{University of Hamburg, Hamburg, Germany}
\affil[4]{LIGO Hanford Observatory, Richland, WA, USA}
\affil[5]{Stanford University, Stanford, CA, USA}
\affil[6]{Università degli Studi dell'Aquila, L'Aquila, Italy}
\affil[7]{Massachusetts Institute of Technology (MIT), Cambridge, MA, USA}
\affil[8]{Vrije Universiteit Amsterdam, Amsterdam, Netherlands}
\affil[9]{Sapienza Università di Roma, Rome, Italy}
\affil[10]{INFN Roma 1, Rome, Italy}
\affil[11]{California Institute of Technology (Caltech), Pasadena, CA, USA}

\date{}

\begin{document}

\maketitle

\begin{abstract}
GEMINI is an underground research and development facility dedicated to advancing seismic isolation and control technologies for future gravitational-wave observatories, including the Einstein Telescope (ET) and the Lunar Gravitational-Wave Antenna (LGWA). This paper presents the technical design and theoretical framework of GEMINI's active seismic isolation platforms, including detailed noise budget analyses, performance predictions, and residual platform motion evaluations. The GEMINI platforms are designed to achieve unprecedented vibration isolation, targeting motion suppression across the 10\,mHz to 10\,Hz frequency band, and with the goal to make them the quietest platforms of their kind. In the context of ET, GEMINI will enable the development and validation of inter-platform control strategies essential for the stabilization of auxiliary degrees of freedom of its interferometers. GEMINI will also support the testing of cryogenic payloads and ultra-sensitive inertial sensors required for LGWA. By integrating advanced cryogenic systems, precision inertial sensors, and state-of-the-art vibration isolation technologies, GEMINI will serve as a versatile testbed for next-generation ground-based gravitational-wave detectors and lunar seismometry missions.
\end{abstract}

\section{Introduction}

The direct detection of gravitational waves (GWs) has opened a new era in observational astronomy and fundamental physics, providing insights into the dynamics of compact object mergers and cosmology. The first and second generation gravitational-wave observatories—LIGO~\cite{AbEA2016b}, Virgo~\cite{AcEA2015}, and KAGRA~\cite{AkEA2018}—have already demonstrated the profound scientific potential of GW detection. However, extending the detection bandwidth to lower frequencies remains a critical challenge for next-generation detectors such as the Einstein Telescope (ET)~\cite{PuEA2010, MaEA2020} and space/planetary-based missions like the Lunar Gravitational-Wave Antenna (LGWA)~\cite{LGWA2021, Ajith_2025}.

Seismic and environmental noise are major limiting factors for terrestrial gravitational-wave detectors, especially below 10\,Hz. Meeting the sensitivity goals of ET low-frequency (LF) detector and LGWA requires innovative seismic isolation, high-performance inertial sensors, and advanced control strategies to suppress tilt-to-horizontal coupling, inter-platform motion, and residual vibrations. GEMINI, located 1.4\,km underground in the Laboratori Nazionali del Gran Sasso (LNGS), is the world’s first R\&D facility dedicated to developing and validating these technologies in the 10\,mHz - 10\,Hz band. Its underground location ensures low environmental noise, enabling the testing of next-generation isolation systems, inter-platform control, and cryogenic inertial sensors essential for ET and LGWA:
\begin{enumerate}
    \item ET: Demonstrate inter-platform control to minimize relative motion between large suspended platforms. This assists the control of auxiliary degrees of freedom (DOFs) in the ET interferometers, including the stabilization of recycling cavities and mode cleaners. By locking the platforms into an optically rigid body, GEMINI aims at enabling the ET-LF length and alignment control for its ambitious low-frequency sensitivity goal.
    \item LGWA: Implement the Moon Emulator, which is an ultra-quiet, cryogenic environment for the testing of seismometers with sensitivities beyond state-of-the-art instruments. The Moon Emulator will reproduce as best as possible the lunar environment inside its permanently shadowed, polar craters, which are the proposed deployment sites of LGWA.
\end{enumerate}

The facility consists of two independently controlled and actively isolated platforms, operating in vacuum. The system implements high-performance sensors including T360 GSN seismometers~\cite{T360GSN}, COBRI interferometers for relative motion sensing~\cite{Gerberding2024COBRI, GerIsl2021}, and a Suspension Platform Interferometer (SPI) for differential inter-platform stabilization~\cite{Koeheal2023, DaEA2012}. Cryogenic capabilities are integrated into one platform to enable sensor testing at temperatures compatible with lunar operation. Advanced control schemes, including SISO (Single-Input Single-Output) and MIMO (Multiple-Input Multiple-Output) strategies, will be implemented in GEMINI to address different operational modes. We call "ET control mode" (ECM) the minimization of absolute and relative platform motion in the 12 translational and rotational rigid-body DOFs. Instead, in the so called "LGWA control mode" (LCM), we prioritize minimizing the error signal for ultra-sensitive inertial sensor testing, where platform motion is not the primary concern.

This paper presents the theoretical framework, technical design, and expected performance of GEMINI’s seismic isolation and control systems. Key contributions include:
\begin{itemize}
    \item Detailed noise budget analysis covering all major sources: seismic ground motion, sensor readout noise (T360, COBRI, SPI), actuator noise, electronics noise, and tilt noise coupling.
    \item Predicted residual platform motion in ECM, highlighting the role of blending filters and feedback control loops.
    \item Implementation and performance of inter-platform control via SPI, demonstrating sub-nanometer and sub-nanorad relative motion stabilization across platforms.
    \item Sensor testing methodology, including the use of Wiener filtering (and common-mode subtraction) techniques to reach LGWA sensor sensitivity requirements.
\end{itemize}

By integrating advanced sensing technologies, high-performance seismic isolation, and precision control, GEMINI provides a versatile testbed for the development of key technologies for next-generation GW observatories and lunar missions.

\section{GEMINI Experimental Setup}
\label{sec:GEMexpsetup}
\begin{figure}[ht!]
    \centering
    \includegraphics[width=11.5cm]{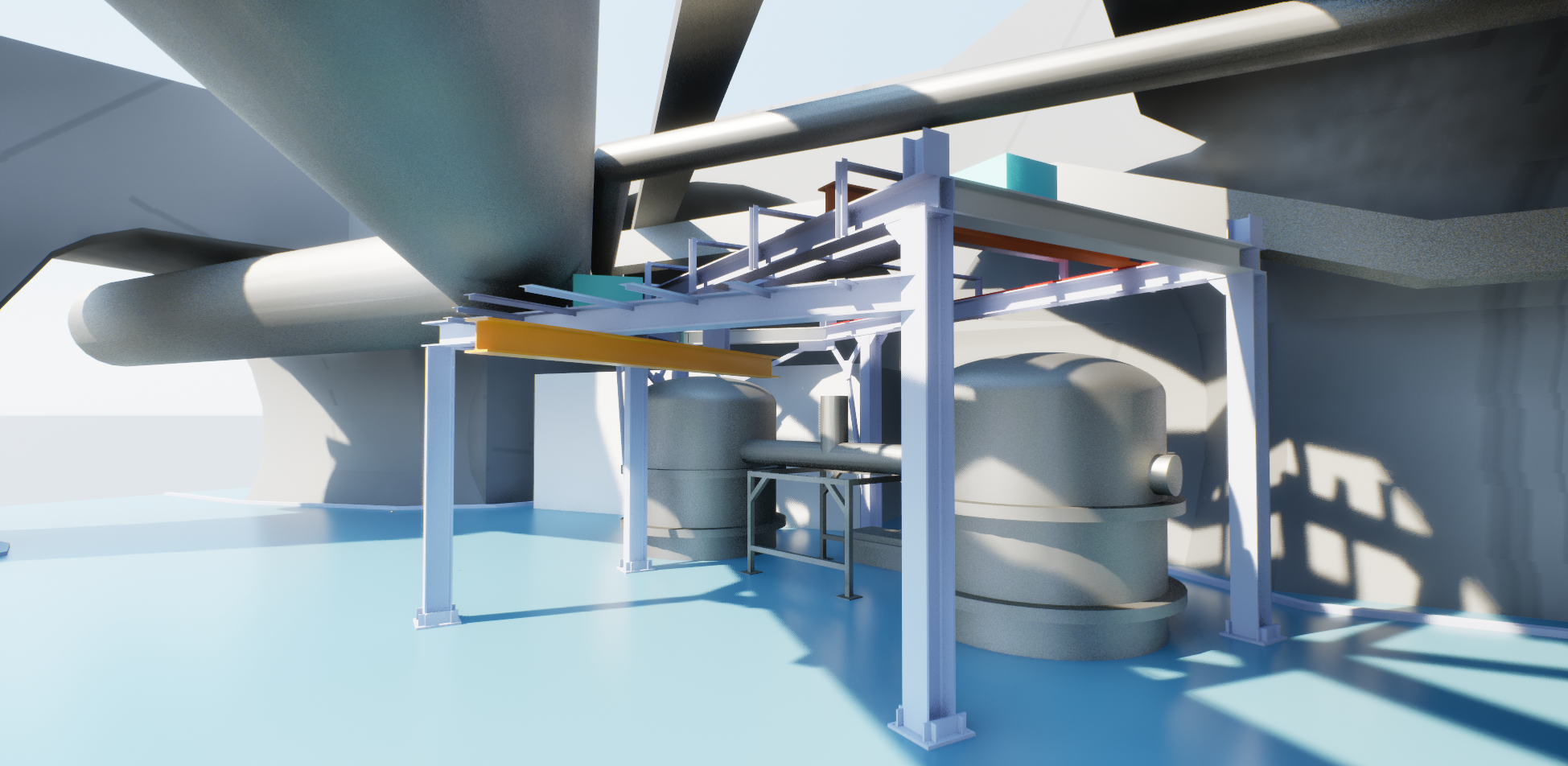}
    \caption{Rendering of the GEMINI site and the cleanroom structural frame. The steel framework surrounds the two experimental chambers and also supports two lifting systems for installation and maintenance. Wall and ceiling panels of the enclosure are omitted for clarity.}
    \label{fig:rendering}
\end{figure}

GEMINI is developed across two laboratory environments: a surface lab used for initial integration and testing, and an underground lab at LNGS, where the final system will be installed. The surface lab accelerates the initial integration work and the development and testing of the control hardware and software. The GEMINI site, located 1.4\,km underground in the Gran Sasso massif, provides a low-noise environment essential for evaluating seismic-isolation and inter-platform control performance. The facility will be designed as a laminar-flow enclosure, equipped with a sufficient number of ceiling-mounted Fan Filter Units (FFUs) to ensure uniform airflow and particle control, targeting ISO Class 6 compliance according to ISO 14644-1~\cite{ISO14644-1-2015}. Figure~\ref{fig:rendering} shows the model of the GEMINI site together with the steel structural framework of the cleanroom, mounted around the two chambers, which will also support two lifting systems required for the installation and maintenance of the experimental components. Note that the wall and ceiling panels of the enclosure are not shown in the figure to provide a clearer visualization.

\subsection{Vacuum System}
\label{sec:vacuum}

\begin{figure}[ht!]
    \centering
    \includegraphics[width=13cm]{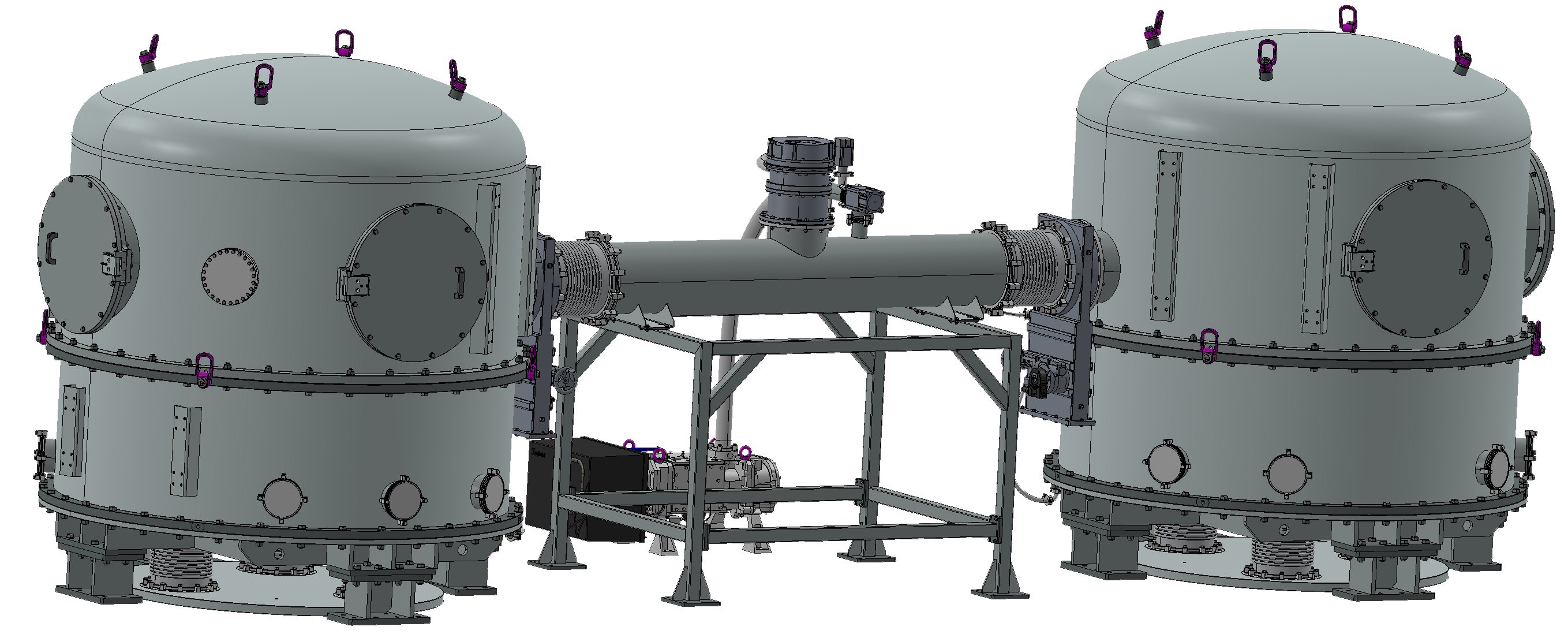}
    \caption{GEMINI vacuum system layout. Two vacuum chambers are connected by a 60\,cm diameter vacuum tube, enabling a laser link for inter-platform measurements.}
    \label{fig:vacuum_sys}
\end{figure}

The GEMINI vacuum system provides an ultra-quiet environment essential for precision displacement measurements and interplatform motion control. It consists of two independent stainless-steel vacuum chambers, each housing an actively isolated platform, and connected by a central vacuum pipe. This pipe enables the laser link required for high-precision interferometric measurements between the platforms. A schematic of the vacuum system is shown in Figure~\ref{fig:vacuum_sys}.

The underground tunnel at LNGS imposes strict dimensional constraints, requiring all components to fit through a 2.2\,m $\times$ 2\,m tunnel entrance. As a result, the vacuum chambers follow a compact, segmented design. Each chamber is constructed from stainless steel (AISI 304L or 316L) and is connected by a 30\,cm diameter vacuum pipe, approximately 3\,m in length, with its central axis positioned 20--30\,cm above the platform plane. Choosing a pipe diameter close to the minimum required for the realization of the inter-platform interferometer was again driven by space constraints, this time at the GEMINI site for the lifting and movement of the vacuum chambers.

The vacuum system is engineered to achieve an ultimate pressure below $10^{-6}$\,mbar, maintained continuously by dry compressing screw and turbomolecular pumps. No bake-out process is currently foreseen. The system is designed to reach the operating pressure within 24 hours of pumping time. Sealing is achieved using metal seals and Viton O-rings.

Both chambers offer convenient access for integration and maintenance. Seven DN150 flanges per chamber provide access to the space below the suspended platforms. The full lateral chamber segment can be lifted from the chamber's base plate for extended lateral access to this space. Three DN600 flanges per chamber with doors enable horizontal access at the level of the suspended optical tables. Three 10-inch diameter flanges with bellows at the base of the chambers enclose the three legs of the platforms. One chamber includes an additional DN200 flange for the cryocooler interface located above the optical table.

The upper dome of each chamber is removable, allowing for the installation of large payloads on the platforms. The internal height of the vacuum chamber accommodates payloads up to 1\,m tall. The system design also accounts for the height limitation imposed by the LNGS hall with a utilizable height of about 3.6\,m. The vacuum chambers are mechanically designed to have their lowest structural resonance frequencies above 15\,Hz. Some parts of the pumping system have lower resonance frequencies. Softer mechanical coupling between the pump units and the vacuum chambers reduces vibration transmission to the isolated platforms. 

\subsection{GEMINI Vibration-Control Platform (GEM-VCP)}
\label{sec:gem-vcp}

The two GEMINI vibration-control platforms (GEM-VCPs) constitute the core mechanical structure for seismic isolation within the GEMINI experiment. Their design is adapted from the LIGO HAM-ISI (Horizontal Access Module - Internal Seismic Isolation) system~\cite{MaEA2014, MaEA2015} with substantial modifications to satisfy GEMINI's stringent low-frequency performance requirements and geometrical constraints imposed by the LNGS underground environment.

\begin{figure}[ht!]
    \centering
    \includegraphics[width=7cm]{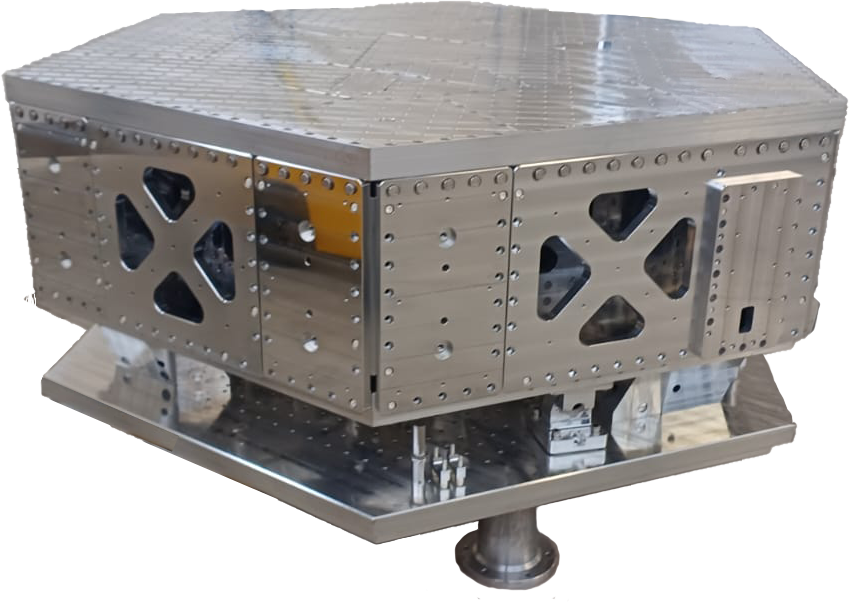}
    \caption{GEMINI's Vibration-Control Platform (GEM-VCP). The two-stage isolation system supports precision instrumentation within the underground vacuum chambers.}
    \label{fig:gem_vcp_photo}
\end{figure}

The GEM-VCPs are two in-vacuum, seismic-isolation platforms separated by a 3\,m baseline. Each platform consists of two stages, stage-0 and stage-1. The stage-0 structure stands on three legs mounted directly onto the laboratory floor, as depicted in Figures~\ref{fig:vacuum_sys} and~\ref{fig:gem_vcp_photo}. The stage-1 structure is a suspended optical table, whose motion is measured and controlled. 

For inertial sensing, GEMINI replaces the GS-13 seismometers used in HAM-ISI with broadband T360 GSN seismometers offering greatly improved vibration measurements below 1\,Hz. The T360 are not vacuum compatible and must therefore be integrated inside vacuum pods. 

A preliminary simplified vibration analysis and structural optimization of the GEM-VCP, including finite-element simulations with ANSYS~\cite{ansys2013}, show that under operational loads the fundamental structural resonance of stage-0 should be above \(70\,\mathrm{Hz}\) and that of stage-1 exceeds \(300\,\mathrm{Hz}\); accordingly, all platform vibration modes lie above \(70\,\mathrm{Hz}\) for stage-0 and above \(300\,\mathrm{Hz}\) for stage-1. These modal frequencies are well above the targeted isolation bandwidth. Anyway, the experimental validation will confirm the analysis data. These modal frequencies are well above the targeted isolation bandwidth (ultimately set by the T360 response function), ensuring that stage-1 structural modes do not interfere with the active control system operating at lower frequencies, that stage-0 does not introduce resonances that could degrade the performance of the suspended stage-1, and that residual deformations at much lower frequencies remain small, thereby justifying the rigid-body platform model used for control. The platform footprint has a minimum diameter of \(1.73\,\mathrm{m}\) at the edges, extending to \(2\,\mathrm{m}\) at the corners. Figure~\ref{fig:gem_vcp_photo} shows the mechanical structure of the GEMINI VCP.

\subsubsection{Stage-0}
\label{sec:stage-0}

Stage-0 forms the structural base of the GEMINI vibration isolation system and is rigidly mounted to the laboratory floor. Its main function is to support the suspended stage-1 without providing active seismic isolation. Unlike the LIGO HAM-ISI design, no inertial sensors are deployed on stage-0. Stage-0 can be treated as a rigid extension of the laboratory floor.

\begin{figure}[ht!]
    \centering
    \includegraphics[width=10cm]{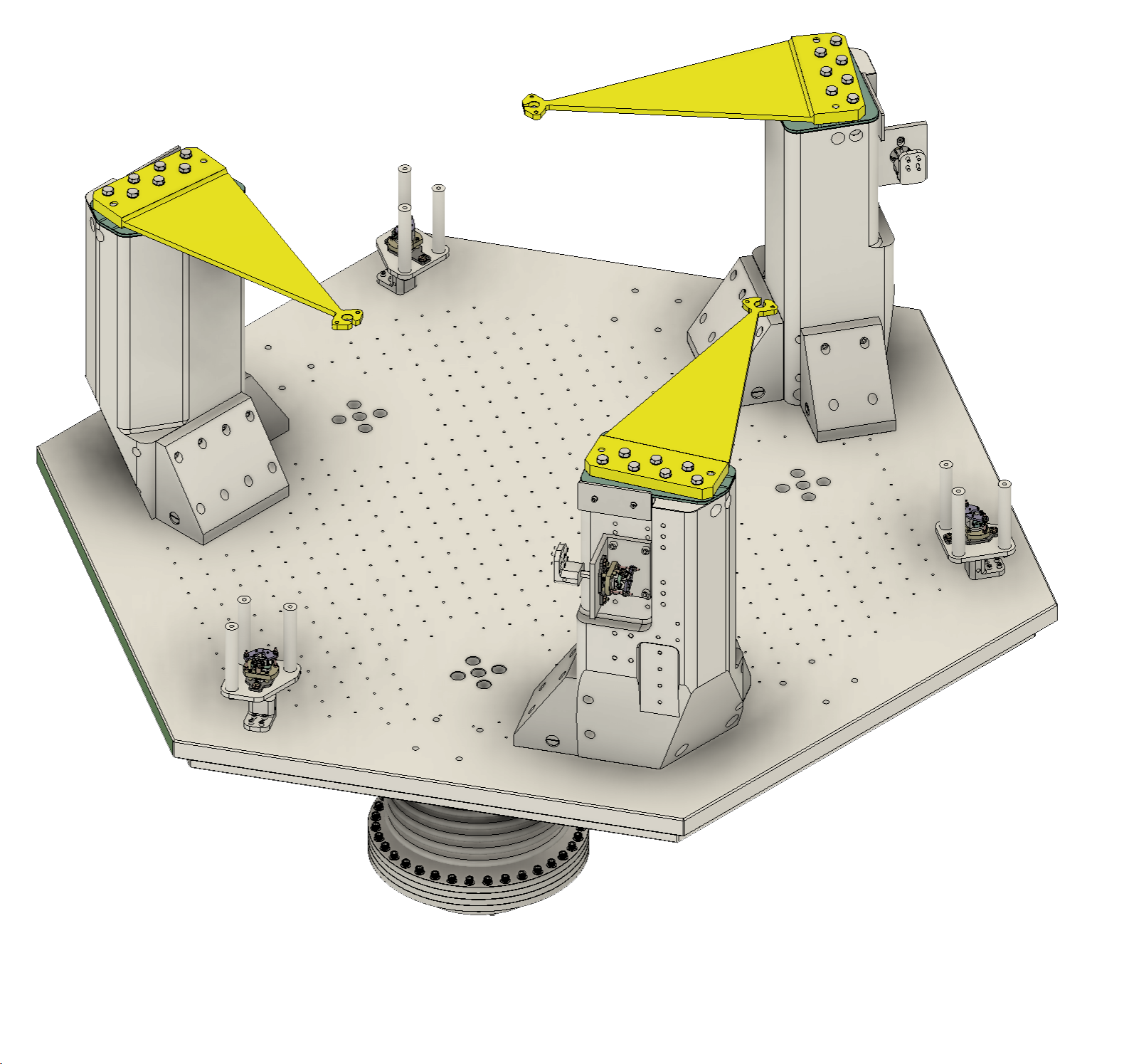}
    \caption{Stage-0 frame, including pillars with yellow indicated spring blades supporting the suspended stage-1.}
    \label{fig:stage-0}
\end{figure}

The mechanical structure of stage-0 consists of a robust aluminum frame, incorporating precision-machined pillars and spring blades, as shown in Figure~\ref{fig:stage-0}. The spring assemblies are arranged to support the suspended stage-1 while minimizing cross-coupling between translational and rotational degrees of freedom. The springs blades designed to be of Titanium Grade 19 (Ti19), ensure that force transmission from stage-0 to stage-1 remains stable and well-controlled, which is critical for maintaining the isolation system’s performance. The exact design of the spring blades will be validated experimentally. All stage-0 components are fabricated from aluminum. Aluminum is compatible with ultra-high-vacuum and, more importantly for GEMINI, an analogous steel structure would have a higher mass.

\subsubsection{Stage-1}
\label{sec:stage-1}
\begin{figure}[ht!]
    \centering
    \includegraphics[width=9.5cm]{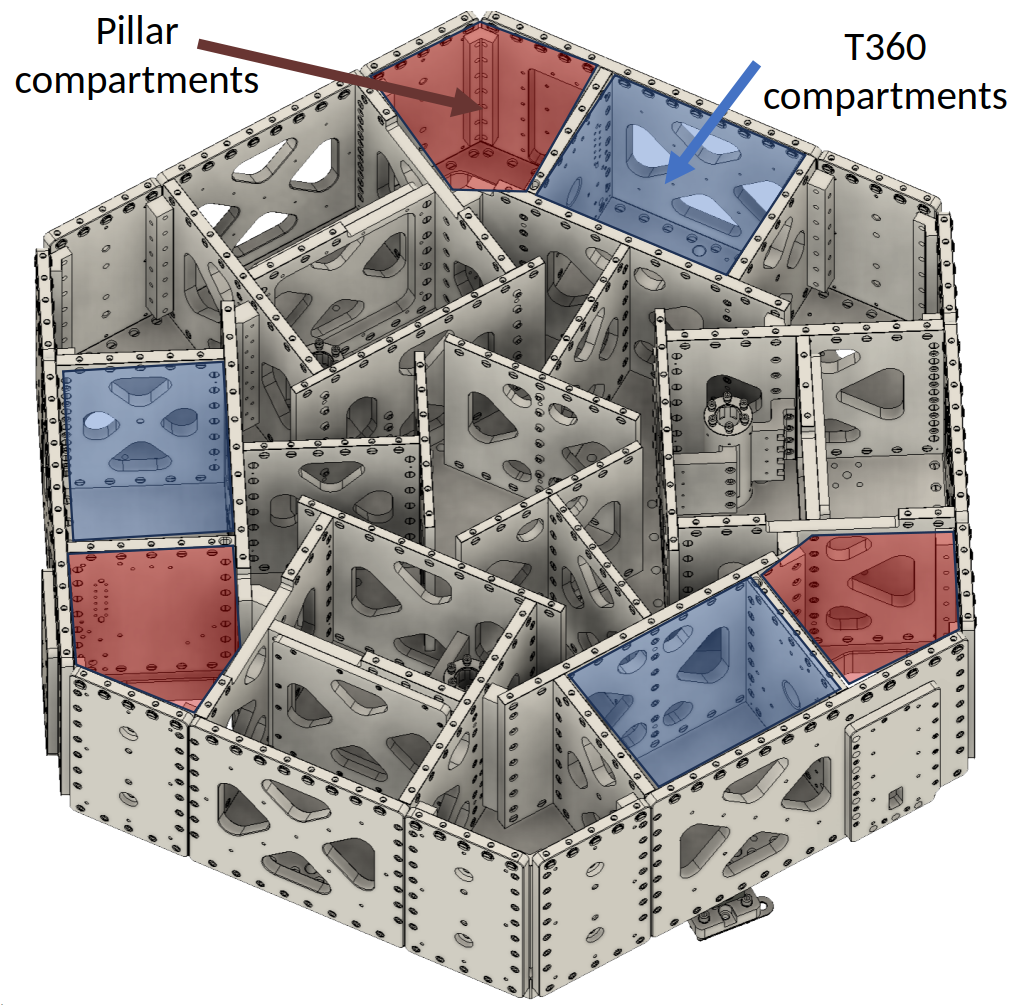}%
    \caption{Stage-1 shown with optical table removed and compartments for T360 GSN sensors indicated.}%
    \label{fig:stage-1}
\end{figure}

Stage-1 is the actively isolated platform of GEMINI suspended from stage-0 by a system of Ti19 spring blades and flexure rods. Inertial sensing on stage-1 is performed by T360 GSN seismometers. These sensors are located inside compartments on stage-1, as illustrated in Figure~\ref{fig:stage-1}. Due to their design, T360 sensors are not compatible with vacuum operation and are therefore placed within custom-designed vacuum pods that provide the necessary environmental control while maintaining compatibility with the chamber’s vacuum requirements. To ensure mechanical protection, mechanical stoppers are mounted between stage-0 and stage-1, limiting their relative motion.

\subsection{Key Components}

GEMINI consists of multiple subsystems designed to achieve the sensitivity required for ET and LGWA technology performance demonstrations. In the following sections, we provide an overview of GEMINI’s primary components and subsystems, including inertial sensors, platform position sensors, the SPI, the actuator system, the environmental monitoring suite, and the real-time control system.

\subsubsection{Sensors}
\label{sec:sensors}

\begin{figure}[ht!]
    \centering
    \includegraphics[width=9cm]{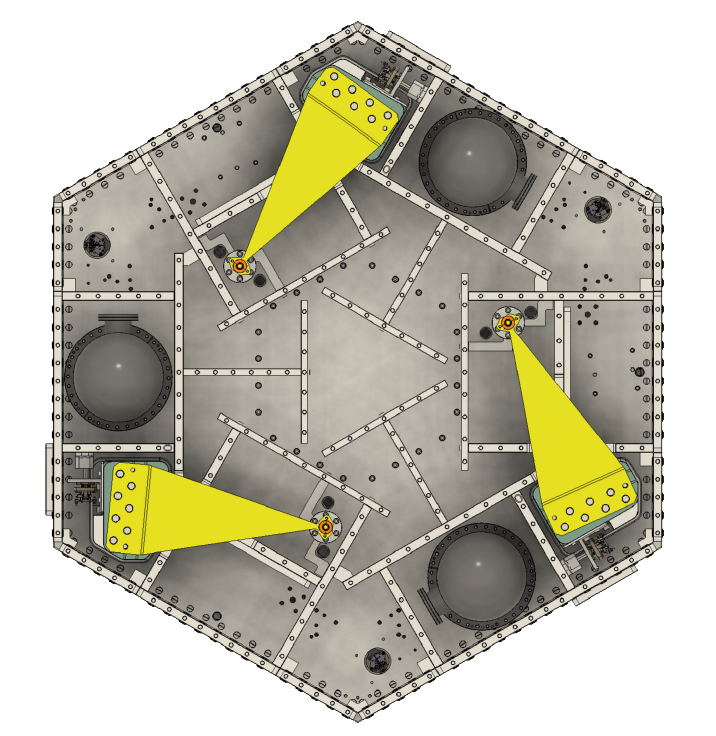}
    \caption{Open stage-1 top view showing vacuum pods housing inertial T360 sensors. The spring-blade-flexure-rod system suspending stage-1 from stage-0 is also visible.}
    \label{fig:stage-1_open}
\end{figure}

Each GEMINI platform is equipped with three high-performance broadband \textit{Nanometrics Trillium 360 GSN Vault} (T360) seismometers. Each sensor provides three-axis inertial measurements with exceptional sensitivity, particularly between 10\,mHz and 50\,Hz. The T360 has an ultra-low magnetic sensitivity of less than 0.03\,(m/s$^2$)/T, which helps to avoid issues with coupling to the magnetic fields produced by the platform's actuator coils. Their deployment within stage-1 is illustrated in Figure~\ref{fig:stage-1_open}. 

\begin{figure}[ht!]
    \centering
    \subfloat[]{{\includegraphics[width=4.5cm]{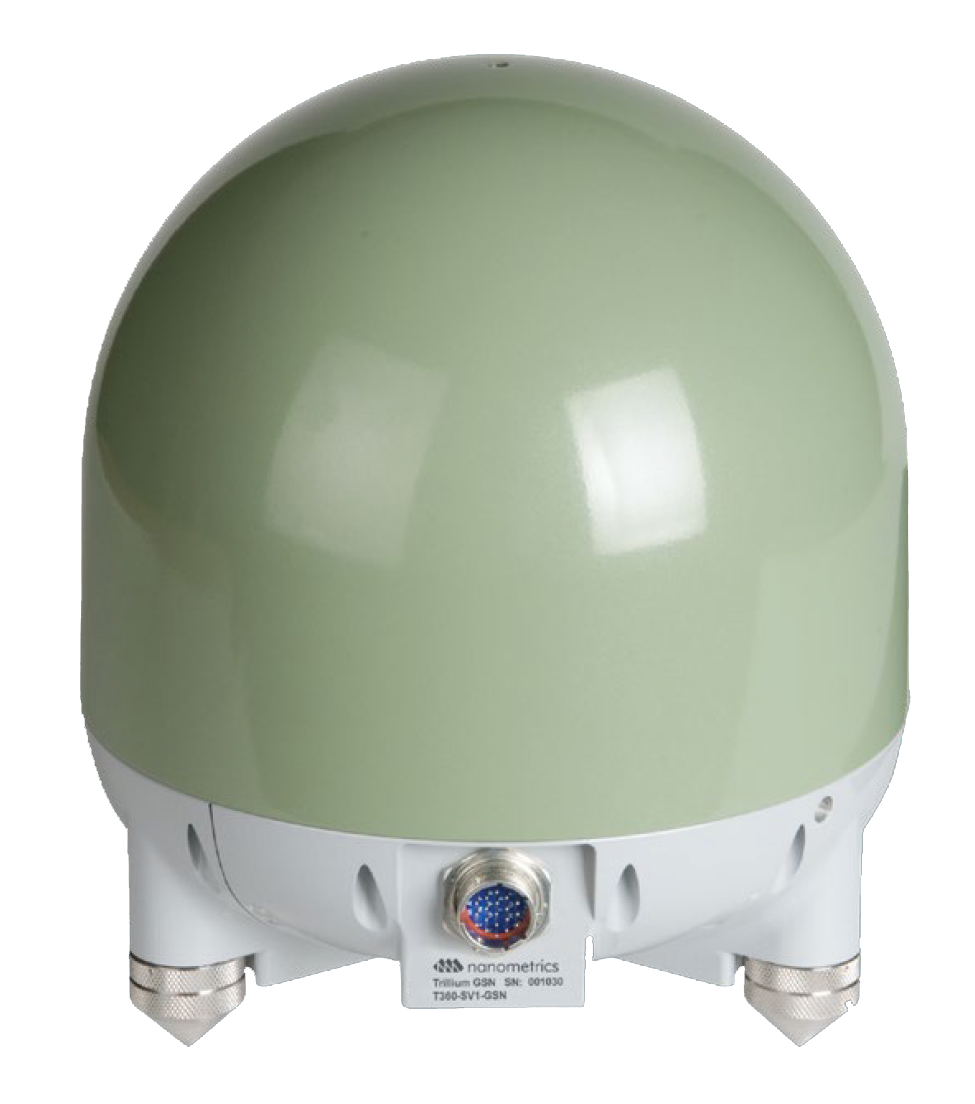} }}%
    \quad
    \subfloat[]{{\includegraphics[width=4.5cm]{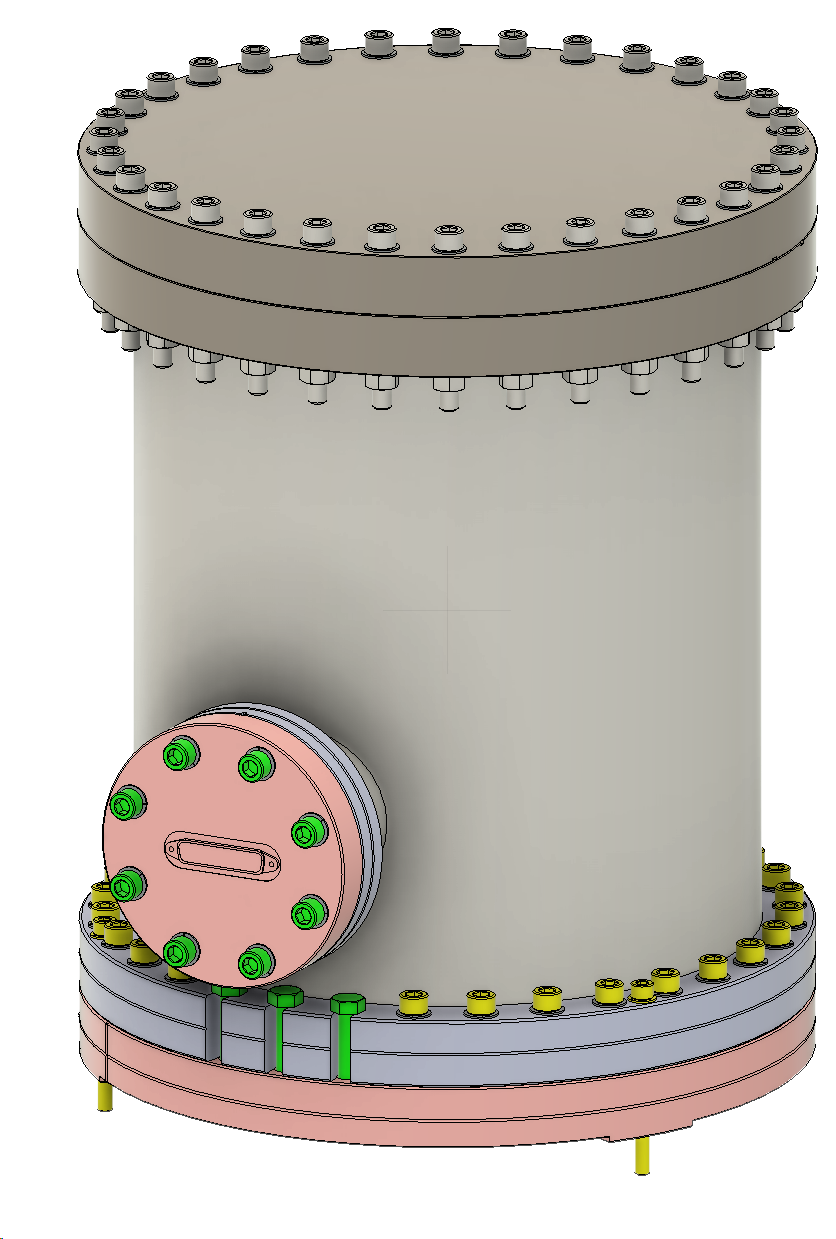} }}%
    \caption{(a) Trillium360 GSN Vault sensor. (b) Vacuum pod enclosing the T360 sensor for operation within the GEMINI vacuum chambers.}
    \label{fig:t360_and_pods}
\end{figure}

Since T360 sensors are not designed to operate under vacuum, they are housed in custom-made vacuum pods (Figure~\ref{fig:t360_and_pods}). These pods are filled with neon gas for leak detection. Their vibrational modes were analyzed to confirm that their natural frequencies lie well above the control bandwidth of the system, ensuring negligible distortion of the measurement of the platform motion.

While the T360 sensors deliver excellent high-frequency inertial sensing, COBRI (COmpact Balanced Readout Interferometer) provides complementary low-frequency displacement sensing \cite{GerIsl2021, Eckhardt2022}. COBRI operates as an inter-stage relative displacement sensor, delivering six degrees-of-freedom measurements of the stage-1 platform’s position and attitude relative to stage-0. The combination of T360 and COBRI sensors allows precise control of platform motion over a wide frequency range.

COBRI utilizes Deep Frequency Modulation Interferometry (DFMI) \cite{gerberding2015deep}, a technique capable of achieving sub-picometer displacement sensitivity over several centimeters of dynamic range. A 3\,GHz sinusoidal phase modulation will be applied to the laser frequency, producing a multi-fringe readout by implementing a real-time readout algorithm \cite{eckhardt2024}. This enables the recovery of displacement information from multiple harmonics of the modulated signal. The COBRI sensor features an on-axis, quasi-monolithic optical layout, providing intrinsic alignment stability when operated in vacuum \cite{GerIsl2021,Gerberding2024COBRI}. Dual readout and balanced detection at the front-end reduce readout noise, suppress scattered light contributions, and mitigate residual amplitude modulation (RAM). The system aims to achieve an effective displacement noise level of approximately 10\,fm/$\sqrt{\text{Hz}}$ between 1\,Hz and 100\,Hz using optical sensing heads that fit into half-inch mirror mounts and have a length of at most 35\,mm.

The SPI measures the relative motion between GEMINI’s two isolated platforms. Its primary function is to synchronize platform motions and enforce an \emph{optically rigid body} (ORB) behavior, where all or at least a relevant subset of the six differential rigid-body degrees of freedom (DOFs) --- displacements and rotations --- are coherently suppressed \cite{HaEA2013}. This suppression of differential motion is essential for ensuring interferometer stability and alignment in ET's vertex region, where the control of auxiliary degrees of freedom, e.g., of the power-recycling and signal-extraction cavities, depend on inter-platform coherence.

The SPI provides high signal-to-noise ratio (SNR) differential displacement and angular measurements. It builds on concepts demonstrated at the AEI 10\,m prototype~\cite{Gosseal2010,Koeheal2023}, where it achieved differential displacement noise levels of 10\,pm/$\sqrt{\text{Hz}}$ at 100\,mHz and angular stability of 1\,nrad/$\sqrt{\text{Hz}}$. However, the SPI in GEMINI must go beyond these benchmarks. Our ambition is to approach displacement sensitivities of approximately 0.1\,pm/$\sqrt{\text{Hz}}$ at 100\,mHz across multiple DOFs, consistent with the stringent requirements of an ORB architecture. Achieving this level of inter-platform coherence across all six DOFs may require a \emph{multi-cavity SPI topology} and potentially even the use of \emph{resonant optical cavities}, as proposed in ORB frameworks~\cite{HaEA2013}. These advanced configurations aim to convert differential displacement in any DOF into detectable longitudinal signals, allowing all relative DOFs to be actively suppressed.

In the context of this paper, GEMINI’s SPI system is tailored for underground platforms and incorporates improved optical lever and interferometric topologies. Several topology options are under evaluation, and each option presents trade-offs between alignment complexity, coupling mechanisms, and achievable sensitivity. A full modeling effort is still required to determine the SPI performance requirements based on the control of auxiliary DOFs at the ET vertex.

\subsubsection{Actuators}
As platform actuators, Sensata Technologies - BEI Kimco's Voice Coil Linear Actuator LA18-32-000A are used. The main characteristics of these actuators are the following: total stroke is 12.7\,mm, continuous stall force is 43\,N, with an actuator constant of 9\,\(\mathrm{N} / \sqrt{\mathrm{W}}\). The total mass of one actuator is 1.3\,kg. It is constructed from low-outgassing materials to make it vacuum compatible. The actuators have a dual-coil design for higher compactness and improved linearity. The design also reduces the strength of the external magnetic field, which is important to suppress magnetic couplings to the inertial seismic sensors on the GEMINI platforms.

\subsubsection{Cryogenic System}

At the core of the cryogenic system is a cryobox cooled to approximately 40\,K. The cryobox emulates the thermal conditions of the Moon’s permanently shadowed craters at its poles, enabling huddle tests of high-precision seismometers in their relevant environment. Cooling is provided by a Sumitomo RDK-500B2 Gifford-McMahon cryocooler. It is capable of providing 45\,W cooling power at 20\,K, with a minimum temperature below 14\,K~\cite{rdk-cryo}. A thermal link connects the cryocooler cold head to the cryobox. The link must be able to extract the heat from the cryobox with an acceptable cool-down time, and at the same time, it must not introduce significant vibrational noise into the VCP. In order to meet both requirements, the main part of the link consists of a suspended copper cylinder whose ends will be connected to the cryocooler's cold head and to the cryobox with short, braided links out of oxygen-free high conductivity copper with a residual resistivity ratio of 150. The detailed geometry of the thermal link has not been decided yet and must also respect integration constraints. The cryobox will be supported by polyetheretherketone (PEEK) cylinders, which have a low thermal conductivity. For radiative insulation, the cryobox will be wrapped in a multi-layer insulation using aluminized Mylar. It has low thermal radiation transfer greatly reducing the heat load on the cryogenic system.

\subsubsection{Environmental Monitoring System}
\label{sec:EMS}
\begin{figure}[ht!]
    \centering
    \includegraphics[width=8cm]{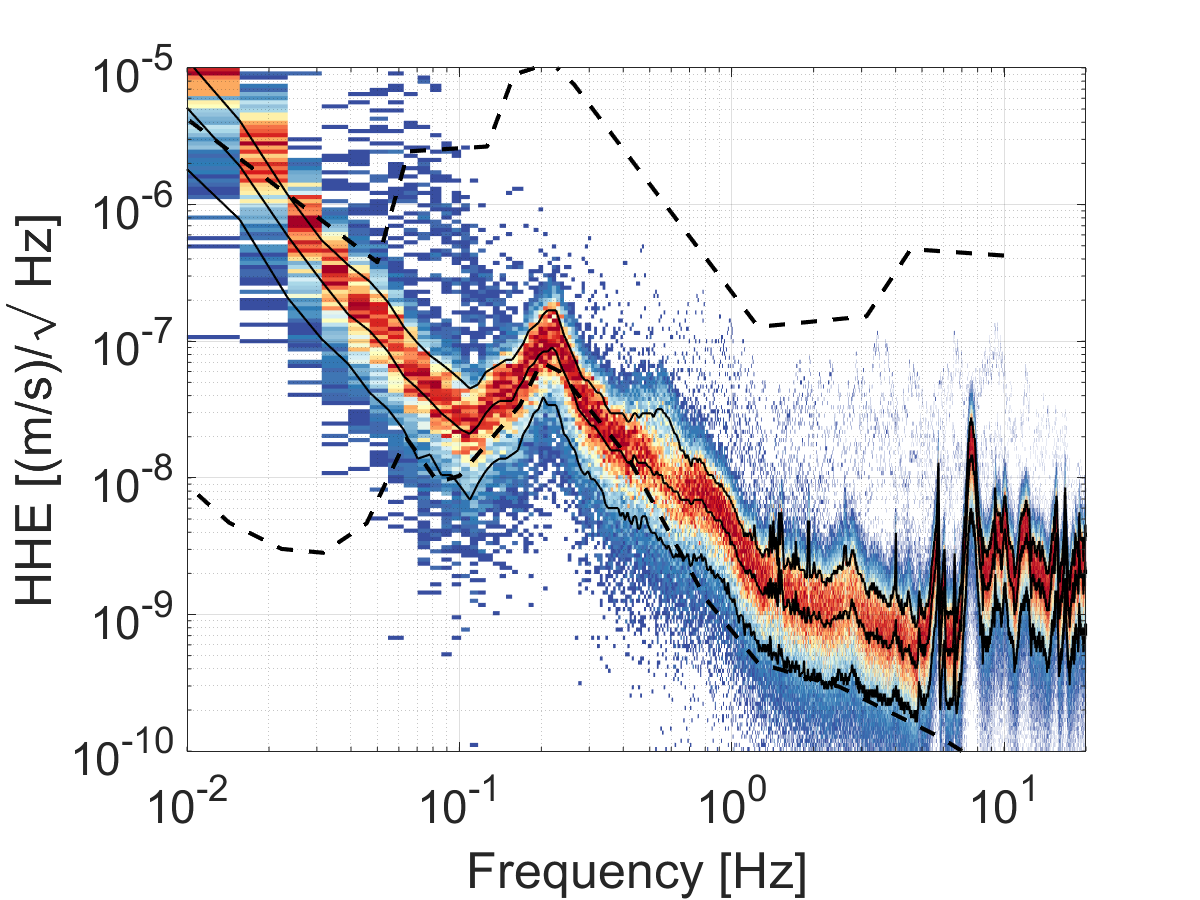}
    \caption{Histogram of spectra of horizontal ground displacement at the GEMINI site. The spectra were calculated from 24 hours of data recorded on July 24, 2025.}
    \label{fig:geminiseismic}
\end{figure}

The environmental monitoring system (EMS) is a critical component of GEMINI, providing continuous data on the ambient conditions at the underground laboratory. A comprehensive understanding of environmental factors such as temperature, pressure, and seismic fields is essential for characterizing and mitigating noise sources that can couple into the VCPs. A 24-hour recording of seismic data at the GEMINI site is shown in Figure \ref{fig:geminiseismic} as a histogram of spectra of horizontal ground displacement. The data were measured with a T360 GSN seismometer connected to a Nanometrics Centaur data logger. An interesting feature only found in horizontal seismometer channels is the excess noise below 0.1\,Hz. It is present in other underground seismic measurements as well and is likely associated with ground tilt. The origin of the tilt is unknown. It could be related to the LNGS active ventilation system or maybe originate from pressure fluctuations at the surface \cite{zuern_2021,washimi_2022}. An array of barometers will be a crucial component of the EMS, and they might potentially be used to correct the horizontal seismic channels to measure with reduced tilt contamination.

\subsubsection{Real-Time System}

The GEMINI experiment relies on a real-time system (RTS) adapted from LIGO’s CyMAC architecture~\cite{CyMACWiki}, with modifications tailored to the specific requirements of GEMINI’s seismic isolation and platform control. The system is responsible for high-speed data acquisition, processing, and actuator driving, forming the backbone of GEMINI’s active isolation and stabilization strategy.

At the core of the RTS is a front-end computer equipped with an X12SPL-F motherboard and an Intel Xeon W-3323 CPU~\cite{supermicro_x12splf}. This configuration provides the computational power necessary to process real-time data from a large number of sensors and to execute complex control algorithms with low latency.

The analog-to-digital converters (ADCs) used in the RTS are the General Standards PCIe-16AI64SSC-64-50M-LIGO (where the LIGO suffix indicates that the card has custom firmware), and the digital-to-analog (DAC) cards are the General Standards PCIe-16AO16-16-F0-DF-LIGO model. Other hardware elements of the RTS include low-noise preamplifiers for input signals, coil-driver modules for the actuators, and whitening filters. Anti-aliasing and anti-imaging filters are incorporated to prevent signal distortion during conversion processes. The timing system ensures precise synchronization of all components, maintaining deterministic behavior in the control loops. The sampling rate of 200\,kHz of the ADCs is not high enough for some of the readout channels. A system of MicroTCA boards is used for the readout and control of the laser-optic system.

\section{Noise Sources and Budget}

GEMINI’s seismic isolation system performance will ultimately be limited by a combination of environmental disturbances, tilt noise, sensor readout noise, and electronics noise in both the sensing and actuator subsystems. This section focuses on the noise budget and the constraints it imposes on the system architecture.

\subsection{Electronics Noise}
\label{sec:electronics_noise_sec}

\begin{figure}
    \centering
    \subfloat[]{{\includegraphics[width=6cm]{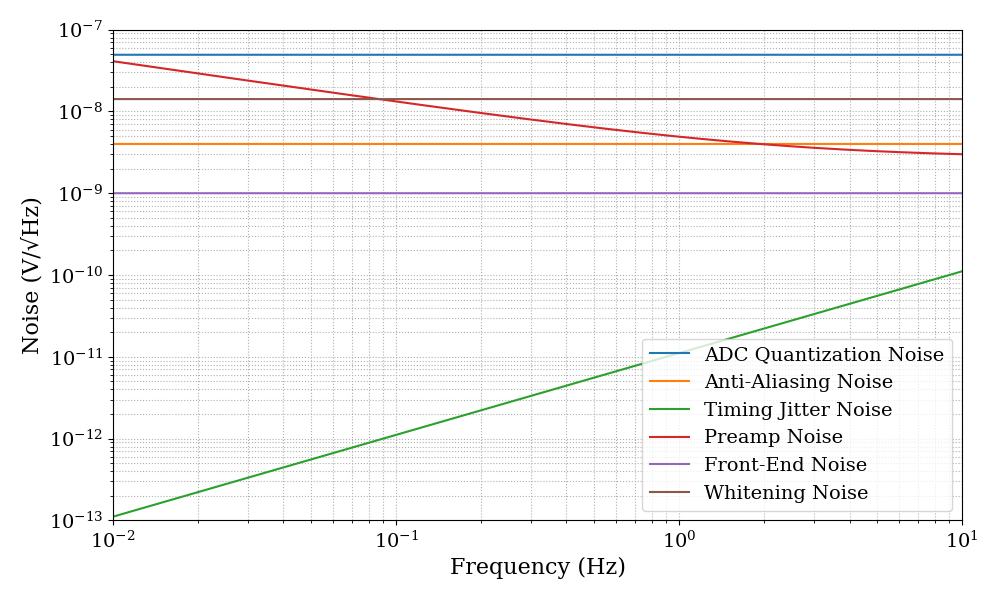} }}%
    \quad
    \subfloat[]{{\includegraphics[width=6cm]{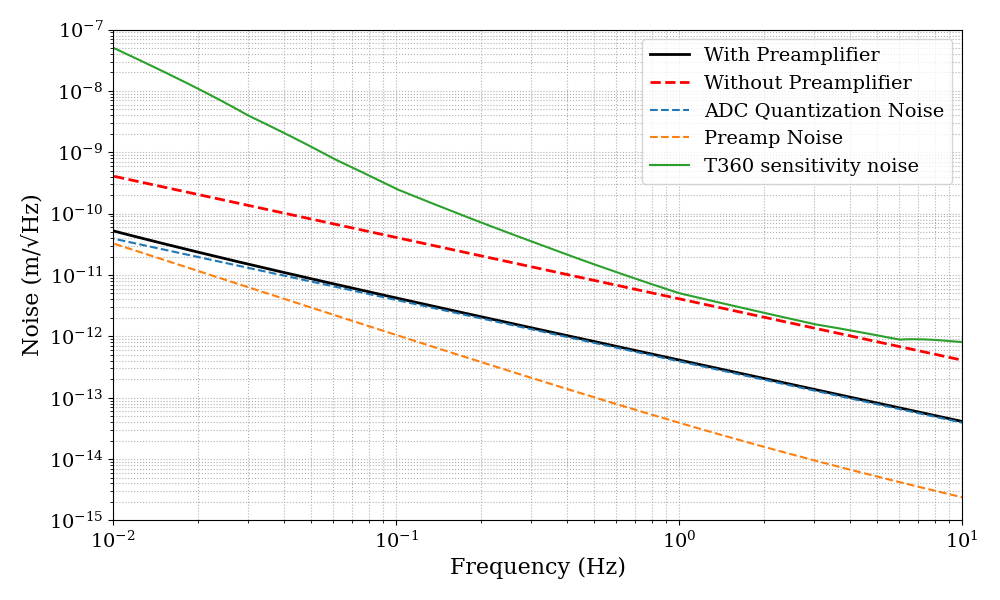} }}%
    \caption{(a) Voltage noise spectral density of individual electronics components. The preamp and ADC quantization noise dominate at low frequencies (b) Equivalent displacement noise of the T360 readout. The system with preamp remains below the seismometer (T360) noise floor over the full 10\,mHz–10\,Hz band.}
    \label{fig:electronics_noise}
\end{figure}

Figure~\ref{fig:electronics_noise}a shows the spectral density of individual electronics noise sources in the voltage domain, while Figure~\ref{fig:electronics_noise}b shows the main noises' effect on the final displacement measurement, together with T360 sensitivity. The following sources are included: ADC quantization noise, preamplifier noise, anti-aliasing filter thermal noise, timing jitter, front-end noise, and whitening amplifier noise.

The ADC introduces quantization noise that depends on its resolution $n_{\mathrm{bits}}$, sampling frequency $f_s$, and input voltage range $V_{\mathrm{range}}$. Its power spectral density (PSD) is given by:
\begin{equation}
    N_{\mathrm{ADC}} = \frac{V_{\mathrm{range}}}{2^{n_{\mathrm{bits}}} \sqrt{12 \cdot f_s}}
\end{equation}
For GEMINI, we employ a 16-bit ADC with a $\pm 2.5$ V input range and $200$\,kHz sampling frequency (then decimated to a lower sampling rate). This yields a quantization noise floor of approximately $50\,\mathrm{nV/\sqrt{Hz}}$. The most demanding readout in terms of sensitivity is of the T360 channels. The seismometers have a response factor of 2000\,V/(m/s), which means that the ADC noise amounts to about $4\,\mathrm{pm/\sqrt{Hz}}$ at 1\,Hz. This is already comparable to the T360 self-noise in that band; nonetheless, we will use a low-noise preamplifier with a gain of 10 to push the ADC contribution comfortably below the sensor noise across the observation band.

The preamplifier itself contributes noise, modeled as frequency-dependent voltage noise based on the LT1125 characteristics~\cite{lt1125datasheet}:
\begin{equation}
    N_{\mathrm{preamp}}(f) = N_{\mathrm{white}} \cdot \sqrt{1 + \frac{f_c}{f}}, \quad N_{\mathrm{white}} = 2.7\,\mathrm{nV}/\sqrt{\mathrm{Hz}}, \quad f_c \approx 2.3\,\mathrm{Hz}
\end{equation}
Other electronics noise sources, which are negligible compared to ADC and preamp noise, are modeled using standard expressions. The anti-aliasing filter contributes thermal noise, modeled as Johnson--Nyquist noise from its input resistor ($e_\mathrm{AA} = \sqrt{4 k_\mathrm{B} T R}$) where we assume $R=1\,\mathrm{k\Omega}$ and $T=290\,\mathrm{K}$. Timing jitter noise, arising from phase-to-amplitude conversion due to a 1\,ps clock instability, is given by $e_\mathrm{jitter}(f) = \pi f \cdot \Delta t_\mathrm{jitter} \cdot V_\mathrm{range} / \sqrt{2}$ and is insignificant at low frequencies. Finally, the front-end and whitening amplifiers add frequency-independent voltage noise, conservatively modeled at 1\,nV/$\sqrt{\mathrm{Hz}}$ and 10\,nV/$\sqrt{\mathrm{Hz}}$ respectively, with the latter scaled by $\sqrt{2}$ for differential readout. These components, shown as voltage spectra in Figure~\ref{fig:electronics_noise}a and as equivalent displacement noise in Figure~\ref{fig:electronics_noise}b, remain subdominant to ADC quantization and preamplifier noise across the entire analysis band.

The digital-to-analog converter (DAC) introduces quantization noise to the actuator command signal. Its spectral density follows the same expression as for the ADC. For GEMINI, the DAC operates with 20-bit resolution and a $\pm10$ V range, with 2048\,Hz sampling frequency. Two additional sources are the Johnson-Nyquist thermal noises from the actuator's internal coil resistance ($R_{\mathrm{coil}}=2.23\,\Omega$) and from the resistive elements of the anti-imaging (reconstruction) filter ($R_{\mathrm{filter}}=1\,\mathrm{k\Omega}$). These voltage noises are summed in quadrature and converted to current noise using the coil impedance $Z(\omega)=R_{\rm tot}+j\omega L_{\rm coil}$. Current noise is then converted to force noise through the actuator force constant $k_t\simeq13.5~\mathrm{N/A}$, consistent with the datasheet actuator constant $K_p=9.03~\mathrm{N}/\sqrt{\mathrm{W}}$ via $k_t=K_p\sqrt{R_{\rm coil}}$. For comparison with other displacement inputs, we optionally form the equivalent pre-loop input displacement
\[
S_{\mathrm{act}}^{\mathrm{eq}}(f)=\bigl|P_{\mathrm{ACT}}(f)\bigr|\,S_F(f),
\]
which naturally exhibits the open-loop plant resonance. For angular DOFs, the force noise is converted to torque using the $0.866$\,m lever arm and propagated with the rotational plant. This clarifies that the DAC noise is filtered by the analog/EMF stages (impedance and $k_t$); the open-loop mechanical response is used only for the optional pre-loop displacement equivalence, not for the closed-loop injection.

\begin{figure}[ht]
    \centering
    \includegraphics[width=8cm]{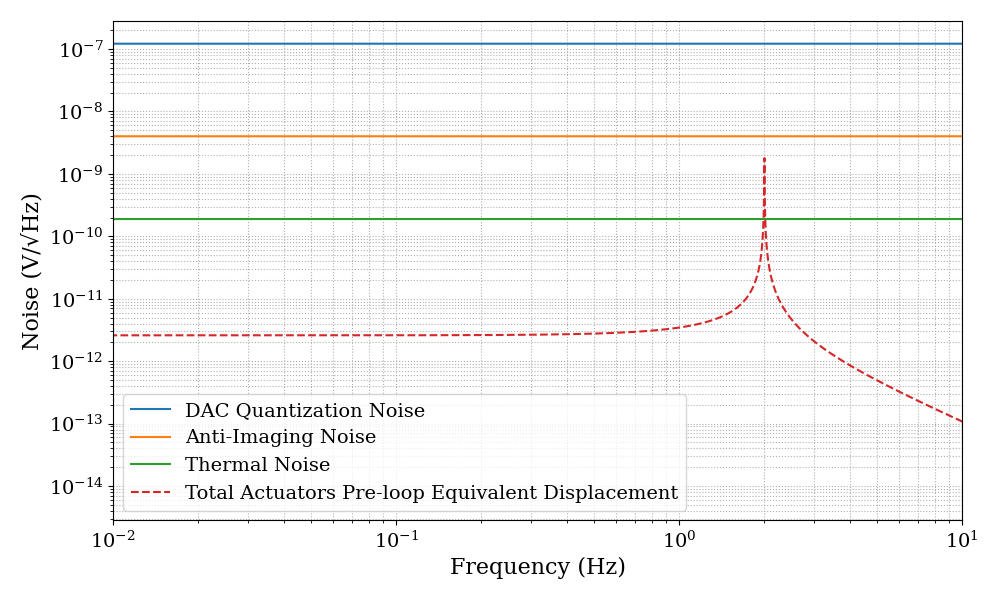}
    \caption{Voltage noise contributions from actuator electronics (quantization, anti-imaging, and thermal), together with the equivalent pre-loop input displacement noise}
    \label{fig:actuator_noise}
\end{figure}

\subsection{Sensor Noises}
The sensor noises determine the performance goals of the GEMINI control loops. The noises of all sensors involved in the VCP are shown as equivalent displacement and rotation noise in Figure~\ref{fig:noise_inputs}. The T360 sensor noise is modeled using the manufacturer-provided self-noise specification, and it sets the baseline sensitivity for the inertial control loop. All three T360 channels have approximately the same self-noise. The magnitude of the T360 response and therefore its sensitivity degrades rapidly below about 3\,mHz and above 60\,Hz. The COBRI position measurements are designed to achieve an effective displacement noise of approximately 10\,fm/$\sqrt{\text{Hz}}$ between 1\,Hz and 100\,Hz, thanks to its DFMI architecture. Despite its low sensor noise, COBRI's contribution to the VCP noise will be significant at low frequencies because of blending filters (see section~\ref{sec:et_mode}). The SPI at the AEI 10\,m prototype has demonstrated a sensitivity of \( 10\,\mathrm{pm}/\sqrt{\mathrm{Hz}}\) at \(0.1\,\mathrm{Hz}\). For GEMINI, we adopt an interim design target of \(0.1\,\mathrm{pm}/\sqrt{\mathrm{Hz}}\) (see Sections~\ref{sec:spi_diff} and~\ref{sec:sensors}). This value is motivated by prior ORB-related studies and will be refined once the control requirements for the auxiliary degrees of freedom at the ET vertices are modeled in detail.

\subsection{Seismic and Tilt Noise}

The seismic input to the GEMINI platforms is based on long-term underground monitoring data from the Gran Sasso INFN Seismic Array (GIGS)~\cite{gigs_data} and has recently been complemented by T360 measurements at the GEMINI site (see section \ref{sec:EMS}). These inputs are used to compute translational and rotational seismic-noise contributions to the platform motion.

\begin{figure}[ht]
    \centering
    \includegraphics[width=1\linewidth]{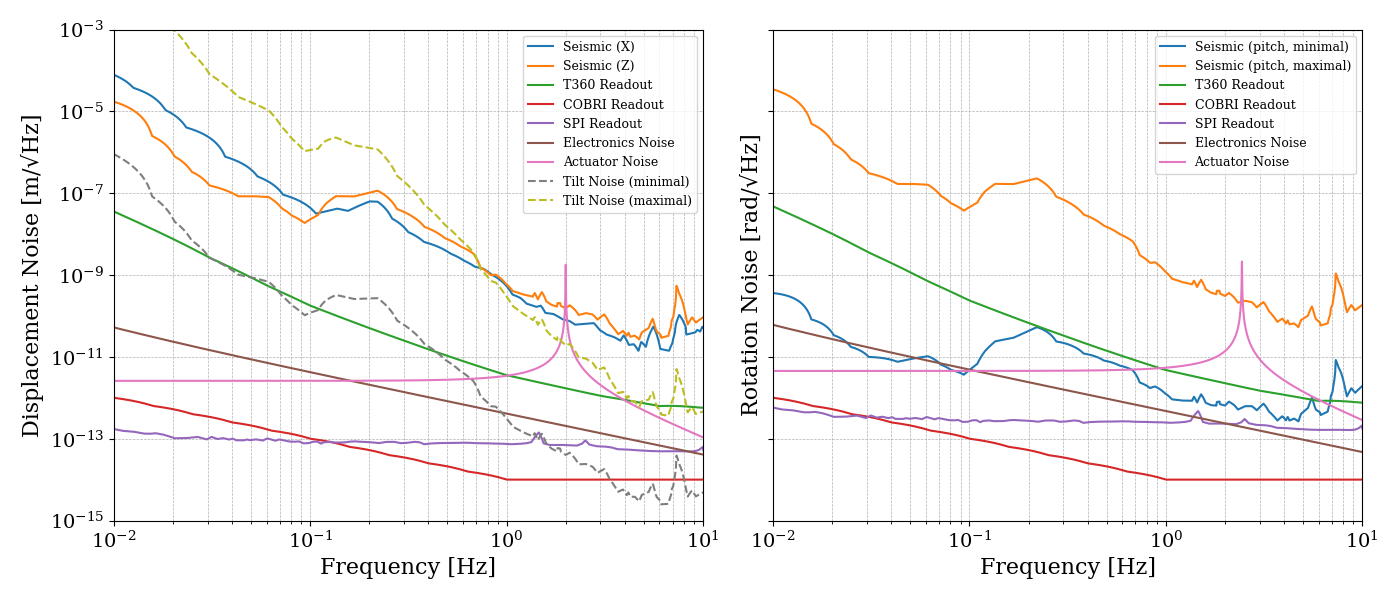}
    \caption{(a) Seismic and readout noise inputs contributing to the longitudinal \(X\)-DOF, including tilt coupling via vertical seismic motion. (b) Pitch noise computed with minimal and maximal coupling to underground vertical seismic motion.}
    \label{fig:noise_inputs}
\end{figure}

Figure~\ref{fig:noise_inputs}a shows the longitudinal noise inputs (X), including vertical seismic motion (Z), along with all relevant readout and actuation noise sources. The vertical component is shown explicitly because it produces platform tilt, which then contributes to the horizontal displacement measurement via tilt-to-horizontal coupling. Figure~\ref{fig:noise_inputs}b shows the rotational seismic input in pitch (rotation about the Y-axis), computed under two scenarios. In the case of minimal coupling, ground tilt is determined by the length of seismic waves traveling with a velocity \(v_{\text{\rm seis}} = 3000\,\mathrm{m/s}\):
\begin{equation}
    \theta_{\text{minimal}} = \frac{2\pi f}{v_{\text{\rm seis}}} N_{\text{\rm seis},Z}
\end{equation}
In the case of maximal coupling, we assume that the full differential ground displacement in the vertical direction acts coherently on the stage-0 legs, which have a distance $D$ to each other,
\begin{equation}
    \theta_{\text{maximal}} = \frac{2}{D} N_{\text{\rm seis},Z}.
\end{equation}
These expressions are used to evaluate the resulting pitch and tilt-induced motion in the longitudinal direction.

The platform tilt contributes a horizontal signal to the T360 measurements, because the seismometer's suspended proof mass is pulled by gravity along its $X$-axis,
\begin{equation}
    N_{\text{tilt}} = \frac{g}{\omega^2} \theta,
\end{equation}
where \(g\) is the gravitational acceleration and \(\omega = 2 \pi f\) is the angular frequency. If the residual tilt motion \(\theta\) is not sufficiently suppressed, it will propagate into the \(X\)-DOF indistinguishable from true horizontal motion without additional tilt sensors.

As we will see later in Figure~\ref{fig:contributions_platform_1} and Figure~\ref{fig:platform_1_residual}, even when the platform tilt is suppressed to the T360 sensing noise level measured differentially by two vertical channels, the tilt contribution to the residual \(X\) motion remains dominant below \(0.6\,\mathrm{Hz}\).

\section{Seismic Isolation}
The GEMINI platforms will be equipped with low-noise instrumentation to ensure precise alignment and inertial isolation across a frequency range spanning from 10\,mHz to 10\,Hz. Seismic isolation in GEMINI consists of both passive and active systems. Passive seismic isolation is achieved through a set of three spring-blade-flexure-rod assemblies, while active isolation employs T360 inertial sensors, the COBRI system, and the SPI, alongside control loops and actuators.

\subsection{Passive Seismic Isolation}
The passive isolation system in GEMINI relies on three spring-blade-flexure-rod assemblies positioned symmetrically at 120-degree intervals around the vertical axis of the suspended platform. If well balanced, the configuration will exhibit small cross-couplings between the six rigid-body degrees of freedom when moving freely or upon actuation. 

\begin{figure}%
    \centering
    \subfloat[]{{\includegraphics[width=11.1cm]{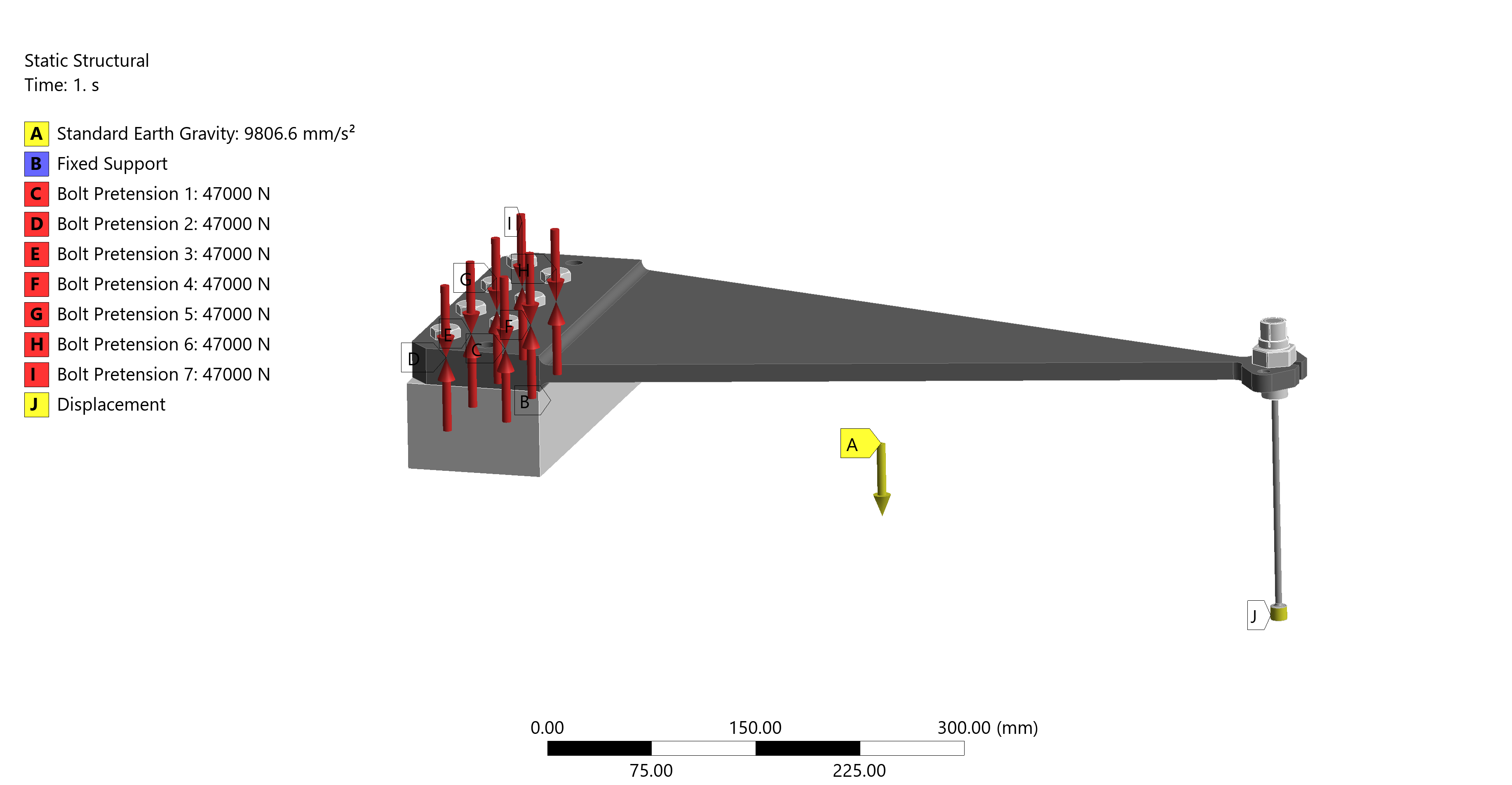} }}%
    \,
    \subfloat[]{{\includegraphics[width=11.1cm]{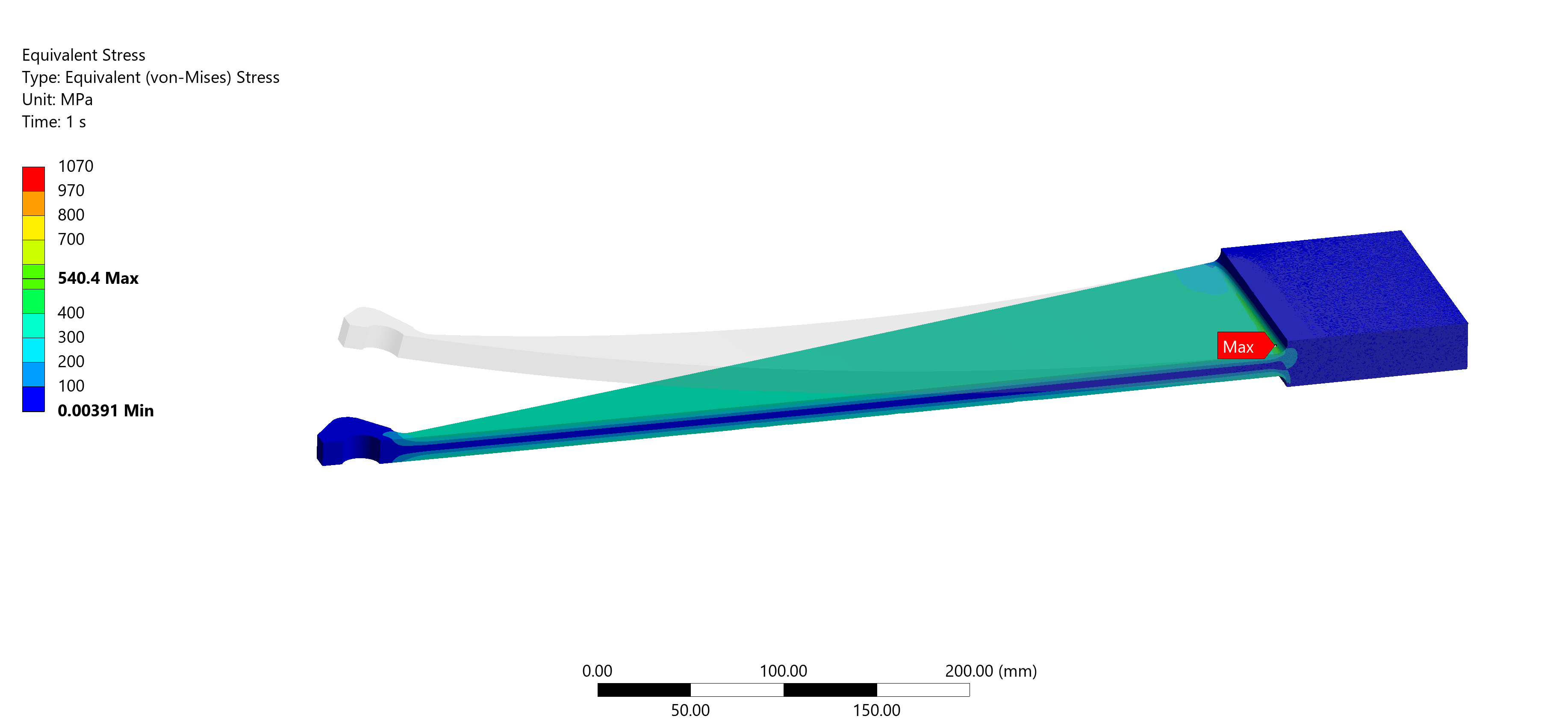} }}%
    \caption{Spring-blade–flexure-rod assembly for GEMINI (a), ANSYS static analysis: equivalent (von-Mises) stress distribution in the Ti-19 spring blade at the nominal operating deflection; the location of the maximum stress is indicated (b).}%
    \label{fig:spring_blade_rod}
\end{figure}

The Ti-19 spring blades are the primary elements responsible for vertical isolation in GEMINI: they are intentionally curved when unloaded and flatten under the suspended load, providing vertical compliance while maintaining structural integrity; the roots are rigidly clamped to minimize unwanted flexibility and frictional hysteresis, and the planform is chosen to meet a target vertical stiffness. The total suspended mass is \(m_t = 1500~\text{kg}\), nominally \(m = m_t/3 = 500~\text{kg}\) per blade to reach the flat-deflection operating point. The mass budget comprises stage-1 (\(800~\text{kg}\)), the optical table (\(255~\text{kg}\)), sensors (\(180~\text{kg}\)), and an adjustment (balance) mass (\(265~\text{kg}\)). This load state is crucial to set the correct working curvature of the blades; the adjustment mass is used to fine-tune the load sharing on each of the three supports so that every blade attains the target deflection and the platform is precisely leveled, compensating for as-built variations in material properties or component masses. The blade is modeled as a trapezoidal bending spring; within Euler–Bernoulli theory the effective vertical stiffness scales approximately as~\cite{Roark2012}
\[
k_z \propto \frac{E\,w\,t^3}{L^3},
\]
where \(E\) is Young’s modulus and \(w\), \(t\), and \(L\) denote the characteristic width, thickness, and length. For GEMINI, \(L=510\,\mathrm{mm}\) with planform width tapering from \(\sim 250\,\mathrm{mm}\) (root) to \(\sim 17\,\mathrm{mm}\) (tip); the current nominal thickness is uniform at \(h=13.2\,\mathrm{mm}\). In the unloaded state, the blade forms an angle of 16.72° from horizontal, which flattens under the operational load. This curvature ensures the desired isolation performance while keeping stress levels within the elastic regime. Each blade connects to stage-1 via a flexure rod designed for horizontal flexibility. The flexure rod has a diameter of 2.99\,mm and a length of 200\,mm. The complete spring-blade--flexure-rod system is shown in Figure~\ref{fig:spring_blade_rod}.

The natural frequencies of the platform are determined by the component properties and the suspended stage-1 mass. The resonant frequencies for the rigid-body modes are expected to be on the order of a few hertz, consistent with the design goals for this type of isolation system. Physically, the horizontal resonance is set by the lateral compliance of the flexure rods, whereas the pitch resonance is set by the much larger vertical stiffness of the load-bearing spring blades. Tilt and vertical resonant frequencies depend on the blade springs' vertical stiffness. For the specific control model and noise budget calculations presented in this paper, representative resonant frequencies of 2\,Hz and 2.44\,Hz are used for the horizontal and tilt degrees of freedom, respectively. The overall conclusions of the noise analysis are not highly sensitive to variations in these frequencies, as the active control system is designed to provide significant suppression in this entire frequency band. The plant transfer functions (TFs) for the seismic paths $P_{\rm GND}$ (displacement to displacement) and the actuator paths $P_{\rm ACT}$ (force to displacement) are given by
\begin{equation}
    P_{\rm GND} = \frac{\omega_0^2 (1 + j \phi)}{\omega_0^2 (1 + j \phi) + s^2} \quad \text{(seismic path plant)},
    \label{eq:seismic_path_plant}
\end{equation}

\begin{equation}
    P_{\rm ACT} = \frac{1}{m \left(\omega_0^2 (1 + j \phi) + s^2\right)} \quad \text{(actuator path plant)},
    \label{eq:actuator_path_plant}
\end{equation}
where $\omega_0$ is the natural frequency of the system, $\phi$ is the loss angle, and $m=1500\,$kg is the mass of the system (the total mass of stage-1 including all of the equipment, has a fixed value).

\subsection{Active Seismic Isolation}
\label{sec:active_seismic}
Active isolation is achieved using inertial sensors (T360 seismometers), which measure the motion of each platform with respect to the inertial reference (the T360 proof mass), the COBRIs, which measure the relative motion between stage-0 and stage-1, and the SPI, which measures the relative motion between the two stage-1 platforms. The signals of the different sensors must be combined for optimal control. For example, the COBRIs are needed for the low-frequency alignment of the platforms to avoid larger excursions of stage-1 with respect to stage-0. This control loop must be blended with the inertial control at higher frequencies using the T360 signals. How to make the best use of the SPI signals is still under investigation. For the noise analyses in this paper, we assume that the SPI signals are used to control one of the two suspended platforms so that it follows the motion of the other one. 

\begin{figure}
    \centering
    \includegraphics[width=13cm]{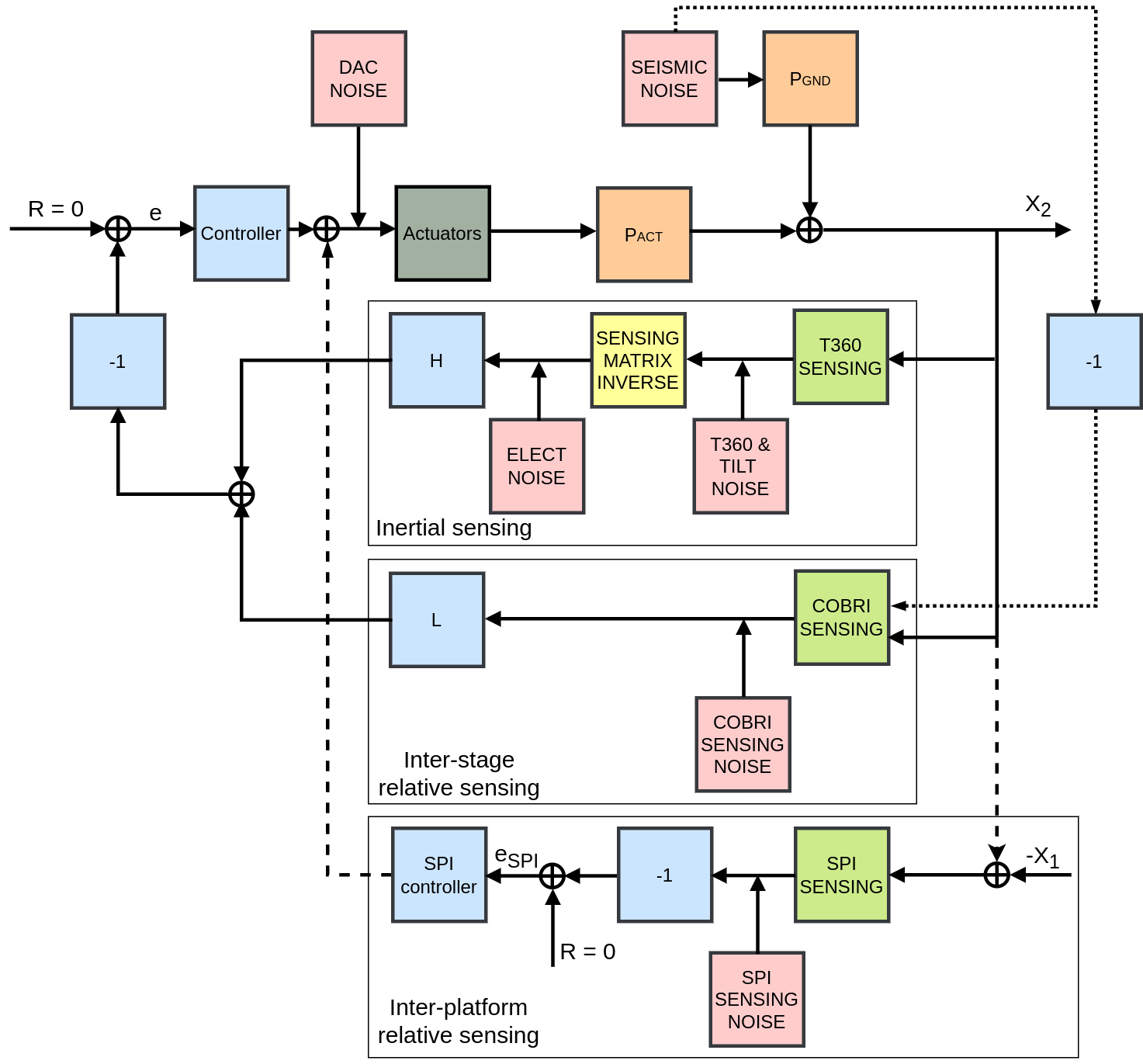}
    \caption{GEMINI's control diagram for a longitudinal DOF. \(P_{\rm GND}\) and \(P_{\rm ACT}\) are the responses of the platform to ground disturbances and actuator forces, respectively. \(X_1\) and \(X_2\) are the platform motions of Platform 1 and 2, respectively. The SPI only contributes to the platform 2 error and control signals. \(H + L = 1\), and the control reference is zero motion.}
    \label{fig:controls_diagram}
\end{figure}

\paragraph{General considerations on the GEMINI control}
Before going to specific applications, we explain first the general idea of how GEMINI's sensing and control works. The inertial sensors measure the displacements X, Y, and Z at three points on stage-1. The T360 signals are transformed to provide the motion of the 6 rigid-body DOFs of each platform (X, Y, Z, pitch, roll, yaw). The redundancy in the horizontal T360 channels can either be used to produce out-of-loop witness channels or to effectively reduce the sensing noise through averaging of signals. As noted earlier, the horizontal channels contain a contribution from the platform tilt via tilt-to-horizontal coupling. The blending of the COBRI and T360 signals is done with a complementary pair of low-pass and high-pass filters, respectively. In our subsequent analysis, the blending frequency is set to 50\,mHz. The outputs of local and SPI measurements are summed up to form the error signal, which is then given to the controller. The controller output goes to the voice-coil actuators. There are three horizontal and three vertical actuators. For example, sending the same signal to all three vertical actuators produces a $Z$-motion of the platform. Differential signals between the vertical actuators produce platform tilt. The magnitude of the platform motion depends on the plant transfer function. This signal is summed with seismic input coming from stage-0 that goes through passive seismic isolation before summation. So we have residual platform motion that again gets sensed inertially and relatively. The flow of signal, noises, plants, controller, and filters are shown in Figure~\ref{fig:controls_diagram}. In this article, in order to determine GEMINI's requirements, we focus on two DOFs, one longitudinal, one angular, \(X\), and pitch. Before discussing the specific control strategies for ET and LGWA modes, Table~\ref{tab:et_mode_symbols} summarizes the symbols, transfer functions, signals, and noise sources used in the subsequent analysis.

\begin{table}[h!]
\centering
\caption{List of symbols used in the ET and LGWA control models.}
\label{tab:et_mode_symbols}
\begin{tabular}{ll}
\toprule
\textbf{Symbol} & \textbf{Description} \\
\midrule
$X_1$, $X_2$ & Platform displacement (Platform 1 and 2) \\
$\theta_1$, $\theta_2$ & Platform pitch rotation \\
$P_{\rm GND}$ & Seismic path plant transfer function \\
$P_{\rm ACT}$ & Actuator path plant transfer function \\
$C_1$, $C_2$ & Local feedback controllers \\
$C_{\mathrm{SPI}}$ & SPI feedback controller \\
$C_{\rm s1}$ & Controller 1 output signal \\
$C_{\rm s2}$ & Controller 2 output signal \\
$H$ & High-pass filter for T360 inertial sensor path \\
$L$ & Low-pass filter for COBRI relative sensor path \\
$U_a$ & Inertial measurement of stage-1 displacement \\
$U_r$ & Stage-0 to stage-1 relative displacement measurement \\
$U_{\mathrm{SPI}}$ & SPI differential measurement \\
$N_{\mathrm{\rm seis}}$ & Ground motion (stage-0 seismic input) \\
$N_{\mathrm{T360}}$ &  T360 sensing noise \\
$N_{\mathrm{tilt}}$ & Tilt-induced noise \\
$N_{\mathrm{ELECT}}$ & Electronics noise \\
$N_{\mathrm{COBRI}}$ & COBRI readout noise \\
$N_{\mathrm{SPI}}$ & SPI readout noise \\
$N_{\mathrm{DAC}}$ & DAC and actuator noise \\
\midrule
\multicolumn{2}{l}{\textit{LGWA Mode Additional Symbols}} \\
$X_{\mathrm{err}}$ & Error signal \\
$C$ & Feedback controller for LGWA mode (single-loop) \\
$U$ & Control output signal \\
$X$ & Residual platform motion in LGWA mode \\
$X_{\mathrm{LGWA}}$ & Output of LGWA sensor \\
$N_{\mathrm{LGWA}}$ & Intrinsic noise of the LGWA science sensor \\
$W(f)$ & Wiener filter \\
$\hat{X}_{\mathrm{LGWA}}$ & Post-filtered LGWA sensor output \\
\bottomrule
\end{tabular}
\end{table}


\subsection{ET Mode: Residual Platform Motion Minimization}
\label{sec:et_mode}
The \textit{ET mode} focuses on minimizing the residual motion of both platforms in \(X\) (horizontal) and \(\theta\) (pitch) DOFs across \(10\,\text{mHz} - 10\,\text{Hz}\), essential in assisting the ET length and alignment control of auxiliary degrees of freedom.

\subsubsection{Residual Motion and Noise Propagation Equations}

The equations expressing the above-described signal and noise flow are given in equation~(\ref{eq:closed_loop_eq_1}) for platform 1. We assume that the input motion and the noises are uncorrelated. Going line by line of the equation~(\ref{eq:closed_loop_eq_1}), we have the stage motion as a function of the input seismic disturbance and the control, then control as a function of the inertial measurements ($U_a$) and the inter-stage measurements ($U_r$) with the COBRI sensors. The readout noises and the tilt contribution are added to these measurements. One of the low-frequency problems is tilt-to-horizontal coupling.

\begin{equation}
    \begin{aligned}
        X_1 &= P_{\rm GND} \cdot N_{\mathrm{seis}} + P_{\rm ACT} \cdot \left( C_{s1} + N_{\mathrm{DAC}} \right) \\
        C_{s1} &= - C_1 \left( H \cdot U_a + L \cdot U_r \right) \\
        U_a &= X_1 + N_{\mathrm{T360}} + N_{\mathrm{tilt}} + N_{\mathrm{ELECT}} \\
        U_r &= (X_1 - N_{\mathrm{seis}}) + N_{\mathrm{COBRI}}
        \label{eq:closed_loop_eq_1}
    \end{aligned}
\end{equation}

In equation (\ref{eq:closed_loop_eq_1_sum}), the PSD $X_1^2$ of the closed-loop motion of platform 1 is given. Going term by term, we see contributions from stage-0 motion ($N_{\rm seis}$), from T360 readout noise ($N_{\rm T360}$), from tilt  ($N_{\rm tilt}$), from electronics ($N_{\rm ELECT}$), of the position sensor noise ($N_{\rm COBRI}$), and of the actuators and DAC ($N_{\rm DAC}$).
\begin{equation}
\begin{aligned}
    X_1^2 &= \left| \frac{P_{\rm GND} + L C_1 P_{\rm ACT}}{1 + C_1 P_{\rm ACT}} \right|^2 N_{\rm seis}^2 \\
    &+ \left| \frac{H C_1 P_{\rm ACT}}{1 + C_1 P_{\rm ACT}} \right|^2 \left( N_{\rm T360}^2 + N_{\rm tilt}^2 + N_{\rm ELECT}^2 \right) \\
    &+ \left| \frac{L C_1 P_{\rm ACT}}{1 + C_1 P_{\rm ACT}} \right|^2 N_{\rm COBRI}^2 \\
    &+ \left| \frac{P_{\rm ACT}}{1 + C_1 P_{\rm ACT}} \right|^2 N_{\rm DAC}^2
    \label{eq:closed_loop_eq_1_sum}
\end{aligned}
\end{equation}

\begin{figure}
    \centering
    \includegraphics[width=13cm]{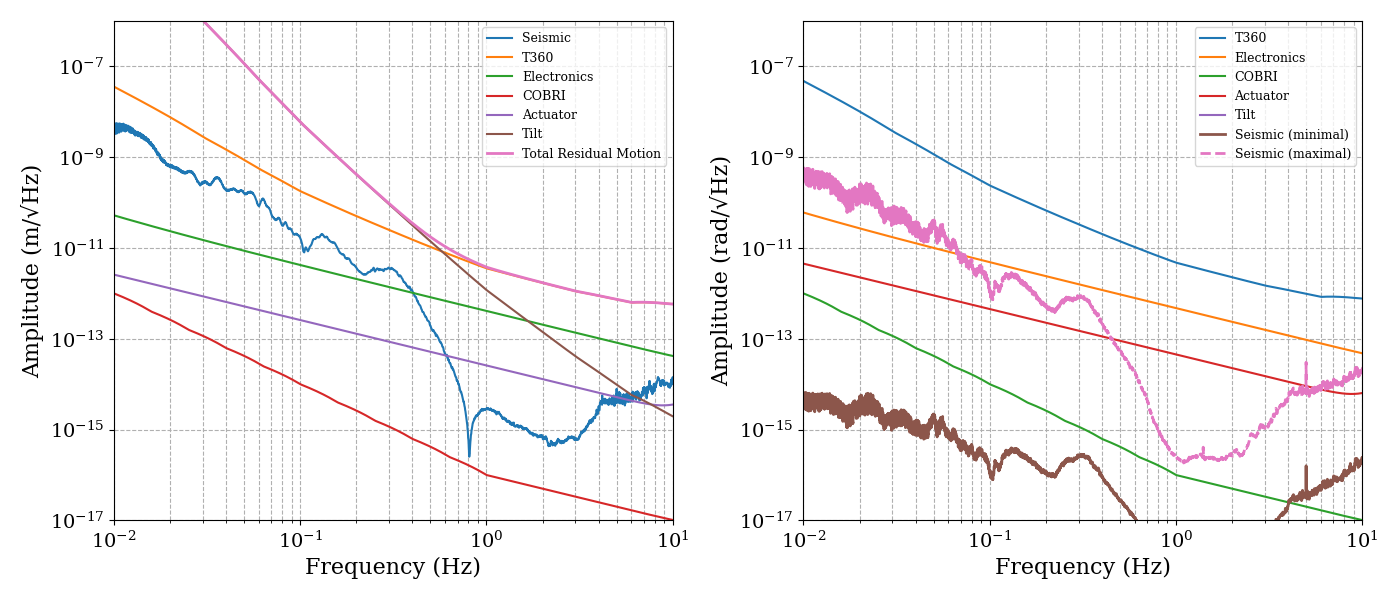}
    \caption{Noise contributions to the closed-loop residual motion of the primary platform for X DOF (left) and pitch DOF (right). The total residual motion of the angular degree of freedom is not shown since it coincides with T360 sensitivity.}
    \label{fig:contributions_platform_1}
\end{figure}

\begin{figure}
    \centering
    \includegraphics[width=13cm]{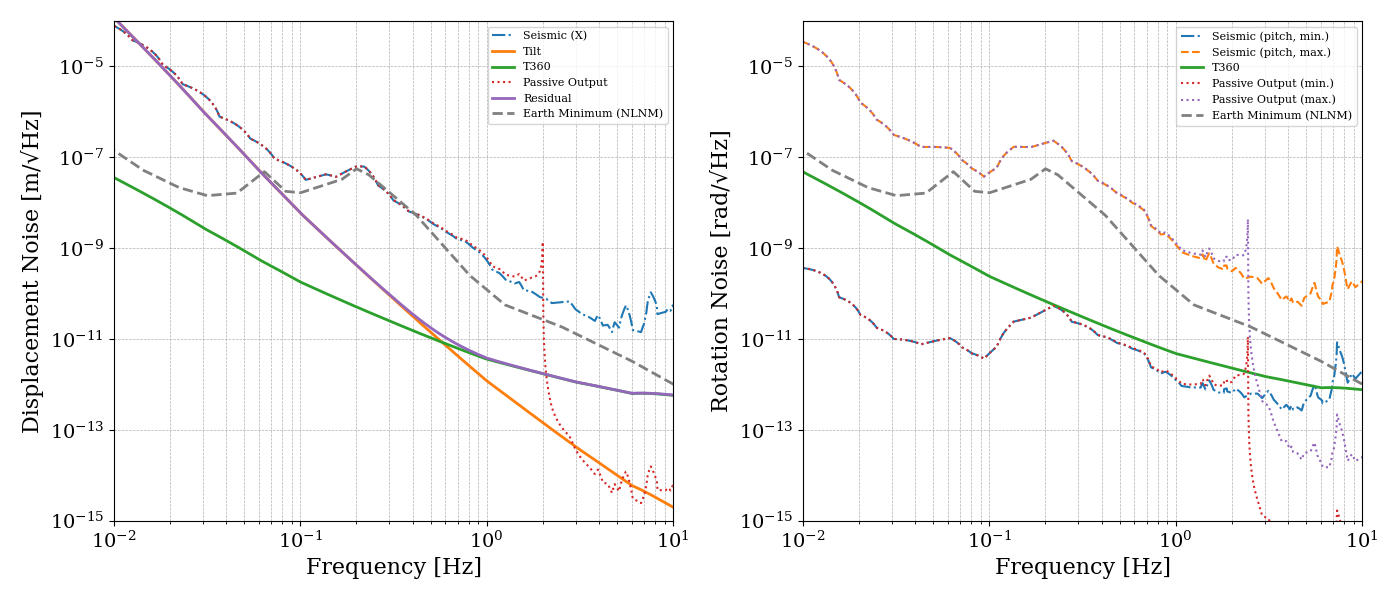}
    \caption{Residual motion of the primary platform in the longitudinal (left) and pitch (right) degrees of freedom. Selected input noise spectra are also shown. For the longitudinal DOF, even when rotational motion is suppressed below the T360 readout noise, tilt-to-horizontal coupling dominates at low frequencies and sets the final residual motion floor in both minimum and maximum-tilt scenarios.}
    \label{fig:platform_1_residual}
\end{figure}

In the case of a high-open loop gain, the final PSD of the closed loop boils down to:
\begin{equation}
\begin{split}
\lim_{C_1 P_{\rm ACT} \to \infty} X_1^2 &= (L N_{\rm seis})^2 + (H N_{\rm T360})^2 + (H N_{\rm tilt})^2 \\
& \qquad + (H N_{\rm ELECT})^2 + (L N_{\rm COBRI})^2
\end{split}
\end{equation}
To clarify, the control architecture uses separate controllers for the longitudinal (\(X\)) and angular (pitch) degrees of freedom. The subscripts in the equations indicate whether the signal pertains to the primary or secondary platform. The residual motion of the primary platform, comprising all contributions as defined in equation~(\ref{eq:closed_loop_eq_1_sum}), is shown in Figure~\ref{fig:contributions_platform_1} with 8th-order control filters. The pitch controller performs effectively across the full frequency band of interest, achieving excellent suppression. For the longitudinal DOF, however, tilt-to-horizontal coupling becomes dominant below 0.6\,Hz, even when the pitch motion is reduced to the sensing noise of a pair of vertical T360 channels. Meeting ET-level residual-motion targets therefore requires an explicit rotational sensor (tiltmeter) and multivariable control to suppress this tilt-to-horizontal contamination (see Section~\ref{sec:tilt_mimo}). The final residual motion spectra, including all noise inputs, is presented in Figure~\ref{fig:platform_1_residual}.

For both the longitudinal (\(X\)) and tilt degrees of freedom, we tuned the feedback filters to yield high loop gain at low frequency and strong attenuation above the mechanical resonances while preserving comfortable stability margins. Stability was verified from the open-loop transfer functions \(C(s)P_{\rm ACT}(s)\) using standard gain/phase-margin tests. The noise-budget results reported in this paper are computed with the corresponding closed-loop transfer functions.

\paragraph{Secondary platform residual motion}
For platform 2 (\textit{secondary}), the control signal includes the SPI feedback:
\begin{equation}
    \begin{aligned}
        X_2 &= P_{\rm GND} \cdot N_{\mathrm{\rm seis}} + P_{\rm ACT} \cdot \left( C_{s2} + N_{\mathrm{DAC}} \right) \\
        C_{s2} &= - C_2 \left( H \cdot U_a + L \cdot U_r \right) - C_{\mathrm{SPI}} \cdot U_{\mathrm{SPI}} \\
        U_a &= X_2 + N_{\mathrm{T360}} + N_{\mathrm{tilt}} + N_{\mathrm{ELECT}} \\
        U_r &= (X_2 - N_{\mathrm{\rm seis}}) + N_{\mathrm{COBRI}} \\
        U_{\mathrm{SPI}} &= (X_2 - X_1) + N_{\mathrm{SPI}}
        \label{eq:closed_loop_eq_2}
    \end{aligned}
\end{equation}

The PSD \(X_2^2\) of the residual motion is given by:
\begin{equation}
\begin{aligned}
    X_2^2 &= \left| \frac{P_{\rm GND} + L C_2 P_{\rm ACT}}{1 + P_{\rm ACT} (C_2 + C_{\rm SPI})} \right|^2 N_{\rm seis}^2 \\
    &+ \left| \frac{H C_2 P_{\rm ACT}}{1 + P_{\rm ACT} (C_2 + C_{\rm SPI})} \right|^2 (N_{\rm T360}^2 + N_{\rm tilt}^2 + N_{\rm ELECT}^2) \\
    &+ \left| \frac{L C_2 P_{\rm ACT}}{1 + P_{\rm ACT} (C_2 + C_{\rm SPI})} \right|^2 N_{\rm COBRI}^2 \\
    &+ \left| \frac{C_{\rm SPI} P_{\rm ACT}}{1 + P_{\rm ACT} (C_2 + C_{\rm SPI})} \right|^2 (X_1^2 + N_{\rm SPI}^2) \\
    &+ \left| \frac{P_{\rm ACT}}{1 + P_{\rm ACT} (C_2 + C_{\rm SPI})} \right|^2 N_{\rm DAC}^2
\end{aligned}
\end{equation}

Due to the symmetric design and control implementation, the residual motion of platform 2 takes a form that closely matches that of platform 1.

In the limit of high open-loop gain, we recover the expected asymptotic expressions:
\begin{align}
\lim_{C_2 P_{\rm ACT} \to \infty} X_2^2 &= (L N_{\rm seis})^2 + (H N_{\rm T360})^2 + (H N_{\rm tilt})^2 \nonumber \\
&\qquad + (H N_{\rm ELECT})^2 + (L N_{\rm COBRI})^2 \\
\lim_{C_{\rm SPI} P_{\rm ACT} \to \infty} X_2^2 &= X_1^2 + N_{\rm SPI}^2
\end{align}


\subsubsection{Differential Motion and the Role of SPI}
\label{sec:spi_diff}

A central objective of the GEMINI control scheme in the ET configuration is to suppress differential motion between the two suspended seismic isolation platforms. This is achieved via the SPI, which measures the relative displacement between the two suspended platforms and provides an error signal used in a high-gain feedback loop with controller \( C_{\mathrm{SPI}} \). The SPI enforces a quasi-rigid-body response between platforms, essential for minimizing differential motion and associated noise coupling into the auxiliary degrees of freedom, e.g., the signal-extraction cavity (SEC), the central Michelson interferometer (MICH), or the power-recycling cavity (PRC).

In frequency-domain simulations, the difference between the residual motion PSDs for platform~1 and platform~2 are minor consistent with the small sensor noise of the SPI. This result is expected given the symmetric configuration: identical controllers, identical plant transfer functions, and uncorrelated noise sources with identical PSDs.

In the residual motion spectrum of platform~2, the contribution from platform~1 enters through the SPI feedback path as 
\[
\left| \frac{C_{\mathrm{SPI}} P_{\rm ACT}}{1 + P_{\rm ACT} (C_2 + C_{\mathrm{SPI}})} \right|^2 X_1^2.
\]
In contrast, the differential motion spectrum includes a cross-term of the form
\[
\left| \frac{1 + C_2 P_{\rm ACT}}{1 + P_{\rm ACT} (C_2 + C_{\mathrm{SPI}})} \right|^2 X_1^2.
\]
All other noise terms—seismic, sensor, electronics, tilt, and actuator—enter the differential motion spectrum identically to how they appear in the platform~2 residual motion.

\begin{figure}[ht]
    \centering
    \includegraphics[width=0.48\textwidth]{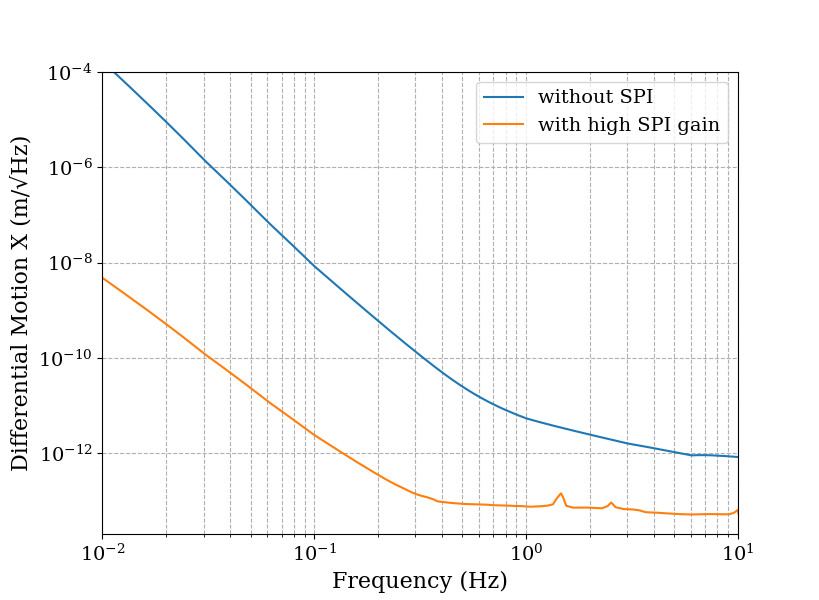}
    \includegraphics[width=0.48\textwidth]{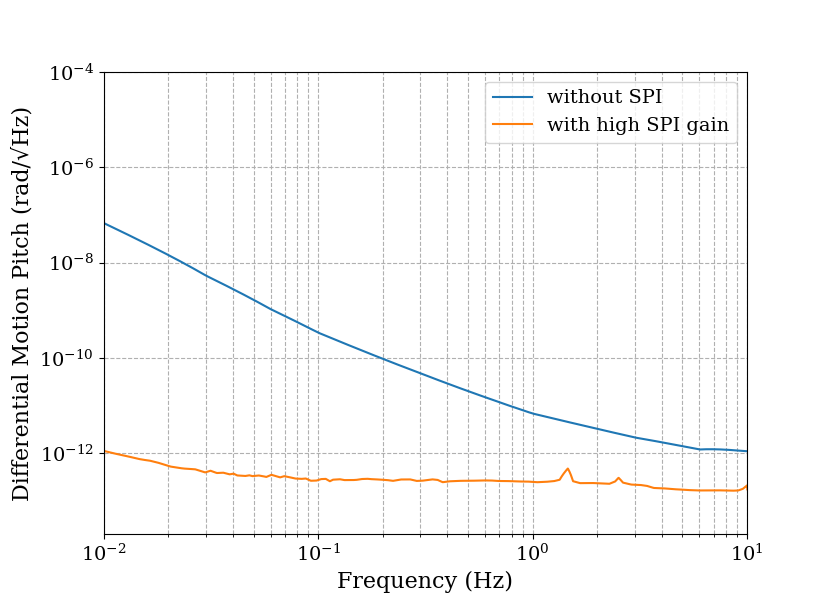}
    \caption{Differential motion spectra in the X (left) and pitch (right) DOFs. Blue curves correspond to the case without SPI, while orange curves represent the case with high SPI gain. SPI enforces inter-platform coherence, leading to significant differential motion suppression, especially at low frequencies.}
    \label{fig:diffmotion_x_pitch}
\end{figure}

In the optimistic case with high SPI gain (stability confirmed), the differential motion between the two platforms is strongly suppressed across the science band, as shown in Figure~\ref{fig:diffmotion_x_pitch}. For the longitudinal ($X$) degree of freedom, the differential motion PSD begins at approximately \(4 \times 10^{-9}\,\mathrm{m}/\sqrt{\mathrm{Hz}}\) at \(10\,\mathrm{mHz}\), decreasing monotonically until it reaches the SPI readout noise floor around \(0.4\,\mathrm{Hz}\), where it flattens out. For the pitch degree of freedom, the differential motion starts at around \(1 \times 10^{-12}\,\mathrm{rad}/\sqrt{\mathrm{Hz}}\) at \(10\,\mathrm{mHz}\), drops further with frequency, and becomes limited by the SPI readout noise above approximately \(0.1\,\mathrm{Hz}\). 

These results highlight the ability of the SPI control system to enforce inter-platform coherence and significantly suppress differential motion, especially at low frequencies where passive residuals could be otherwise dominant. In the absence of SPI, the differential motion spectrum simply reflects the incoherent residual motion of each platform and is approximately larger by a factor of \(\sqrt{2}\). The observed suppression confirms the effectiveness of the SPI system in achieving sub-nanometer and sub-nrad differential stability across the relevant frequency band.

\subsubsection{Tilt-to-Horizontal Coupling and the Necessity of a Tilt-Meter and MIMO Control}
\label{sec:tilt_mimo}

\begin{figure}[h]
    \centering
    \includegraphics[width=10cm]{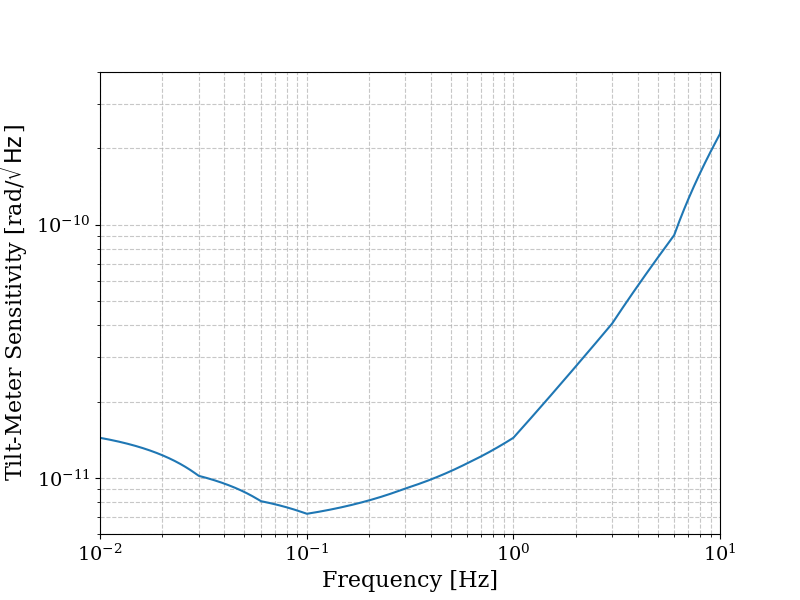}
    \caption{Required tiltmeter sensitivity to fully eliminate tilt contamination in the control of the horizontal platform displacement.}
    \label{fig:tilt_meter_sensitivity}
\end{figure}

At low frequencies, ground tilt contaminates the measurement of the horizontal translational (\(X\)) motion through tilt-to-horizontal coupling. As we have seen in the previous sections, suppressing tilt-induced contamination is essential to meet the stringent residual motion requirements of the ET, particularly in the low-frequency band. To mitigate tilt-induced contamination in \(X\), it is necessary to measure platform tilt with high precision. This makes it possible to distinguish tilt-induced horizontal accelerations from true horizontal motion. The required sensitivity for such a tiltmeter to fully eliminate tilt-contamination in GEMINI is shown in Figure~\ref{fig:tilt_meter_sensitivity}. 

A MIMO controller is required to have coordinated control of both translational (\(X\)) and rotational (\(\theta\)) DOFs. The MIMO control strategy enables simultaneous suppression of both motions by accounting for the cross-coupling terms inherent in the sensing and actuation matrices. 


\subsection{LGWA Mode: Error Signal Minimization}
\label{sec:lgwa_mode}

The LGWA control mode of GEMINI focuses on minimizing the \textit{error signal} \( X_{\mathrm{err}} \), rather than the platform’s residual motion \( X \), to create an ultra-quiet inertial reference frame for testing highly sensitive seismometers intended for LGWA. In the LGWA mode, the control problem concerns a single platform.

The error signal $X_{\rm err}$ contains contributions from inertial (T360) and position (COBRI) sensors:
\begin{equation}
    X_{\rm err} = H \cdot \left( X + \frac{g}{\omega^2} \theta + N_{\mathrm{T360}} + N_{\mathrm{ELECT}} \right) + L \cdot \left( (X - N_{\mathrm{\rm seis}}) + N_{\mathrm{COBRI}} \right).
\end{equation}
Here is $X$, the horizontal motion of the platform, and $\theta$ is the platform's tilt contribution from pitch.

A feedback controller $C$ generates a control signal based on the error signal $X_{\rm err}$, which then drives the actuators producing a force in the horizontal direction:
\begin{equation}
U = - C \cdot X_{\rm err}
\end{equation}

The horizontal platform displacement can then be written as:
\begin{equation}
    X = P_{\rm GND} \cdot N_{\mathrm{\rm seis}} + P_{\rm ACT} \cdot (U + N_{\mathrm{DAC}}),
\label{eq:residual_lgwa_1}
\end{equation}
where \(N_{\mathrm{DAC}}\) represents the actuator and DAC noise. Substituting \(U\):
\begin{equation}
    X = P_{\rm GND} N_{\mathrm{\rm seis}} - C P_{\rm ACT} X_{\rm err} + P_{\rm ACT} N_{\mathrm{DAC}}.
    \label{eq:residual_lgwa_2}
\end{equation}

Inserting the expression for $X$, grouping terms and keeping in mind that $H + L = 1$, we obtain the closed-loop error signal:
\begin{equation}
\begin{split}
    X_{\rm err} &= \frac{1}{1 + CP_{\rm ACT}}\bigg[H \left( P_{\rm GND} N_{\mathrm{\rm seis}} + \dfrac{g}{\omega^2} \theta + N_{\mathrm{T360}} + N_{\mathrm{ELECT}} \right) \\
    &\qquad + L \left( (P_{\rm GND} - 1) N_{\mathrm{\rm seis}} + N_{\mathrm{COBRI}} \right) + P_{\rm ACT} N_{\mathrm{DAC}}\bigg].
\end{split}
\end{equation}

Error-signal suppression in the LGWA mode occurs through the closed-loop gain factor \(1 + C \cdot P_{\rm ACT}\), which attenuates all disturbances, including ground motion \(N_{\mathrm{\rm seis}}\), tilt \(\theta\), and sensor noises. Tilt-induced signals enter the error point similarly to the T360 sensor noise and are suppressed above the blending frequency. In the high control gain limit (\(C \to \infty\)),

\[
X_{\mathrm{err}} \approx 0
\]

This implies that the feedback loop minimizes the error signal by compensating for disturbances through platform motion \(X\), including those arising from tilt. As a result, while tilt motion \(\theta\) remains physically present, its influence on the error signal is canceled, and there is no need for direct tilt sensing, control, or torque actuation.

Even though the residual platform motion $X$ is not directly relevant here, we can derive it, since it will be useful for the next section. Substituting $X_{\rm err}$ in equation~(\ref{eq:residual_lgwa_2}):
\begin{equation}
    \begin{split}
    X = P_{\rm GND} \cdot N_{\mathrm{\rm seis}} 
    - \frac{C \cdot P_{\rm ACT}}{1 + C \cdot P_{\rm ACT}} \bigg[ 
    & H \left( P_{\rm GND} \cdot N_{\mathrm{\rm seis}} + \dfrac{g}{\omega^2} \theta + N_{\mathrm{T360}} + N_{\mathrm{ELECT}} \right) \\
    & + L \left( (P_{\rm GND} - 1) N_{\mathrm{\rm seis}} + N_{\mathrm{COBRI}} \right) 
    + P_{\rm ACT} \cdot N_{\mathrm{DAC}} \bigg] \\
    & +  P_{\rm ACT} N_{\mathrm{DAC}}.
    \end{split}
    \label{eq:residual_LGWA}
\end{equation}

To summarize, the relevant channel in the LGWA mode is the error signal \( X_{\mathrm{err}} \), not the platform’s physical displacement. Tilt sensing or control is not required: the feedback loop intrinsically suppresses tilt noise in the error signal without additional rotational sensing.

\paragraph{Error Signal PSD Calculation}
To evaluate the performance of the control system, we analyze the PSD of the error signal, which quantifies the quality of the inertial reference experienced by the to be tested LGWA sensor. The PSD of $X_{\rm err}(f)$ is computed by summing the contributions of each uncorrelated noise source, weighted by their corresponding transfer functions:
\begin{equation}
    \begin{aligned}
    |X_{\rm err}(f)|^2 &= \left| \frac{P_{\rm GND} - L}{1 + C P_{\rm ACT}} \right|^2 N_{\mathrm{\rm seis}}^2 \\
    &+ \left| \frac{H}{1 + C P_{\rm ACT}} \right|^2 \left( N_{\mathrm{T360}}^2 + N_{\mathrm{tilt}}^2 + N_{\mathrm{ELECT}}^2 \right) \\
    &+ \left| \frac{L}{1 + C P_{\rm ACT}} \right|^2 N_{\mathrm{COBRI}}^2 \\
    &+ \left| \frac{P_{\rm ACT}}{1 + C P_{\rm ACT}} \right|^2 N_{\mathrm{DAC}}^2
    \end{aligned}
    \label{eq:contributions_error}
\end{equation}

\begin{figure}%
    \centering
    \subfloat[]{{\includegraphics[width=11cm]{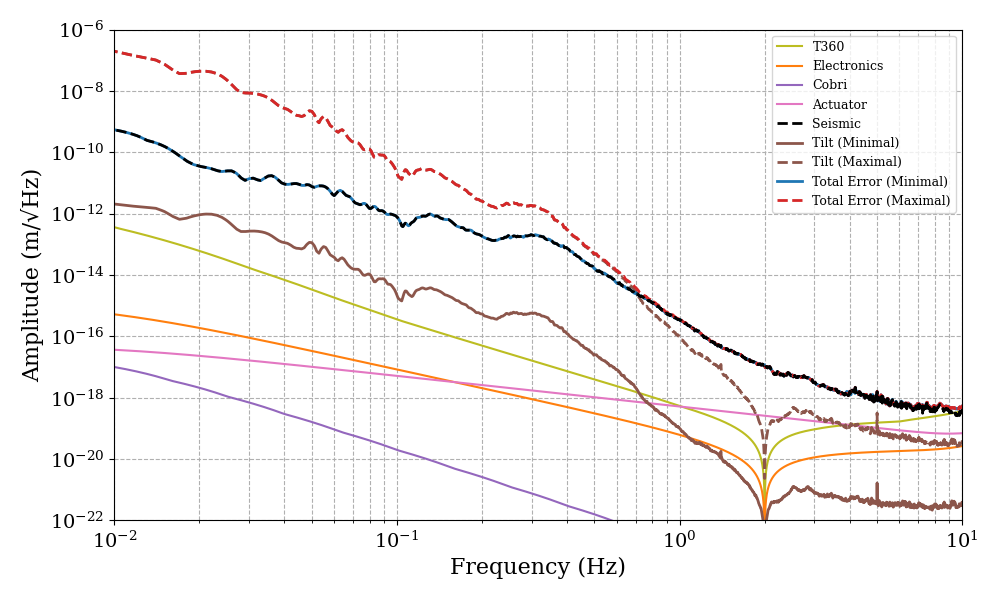} }}%
    \,
    \subfloat[]{{\includegraphics[width=11cm]{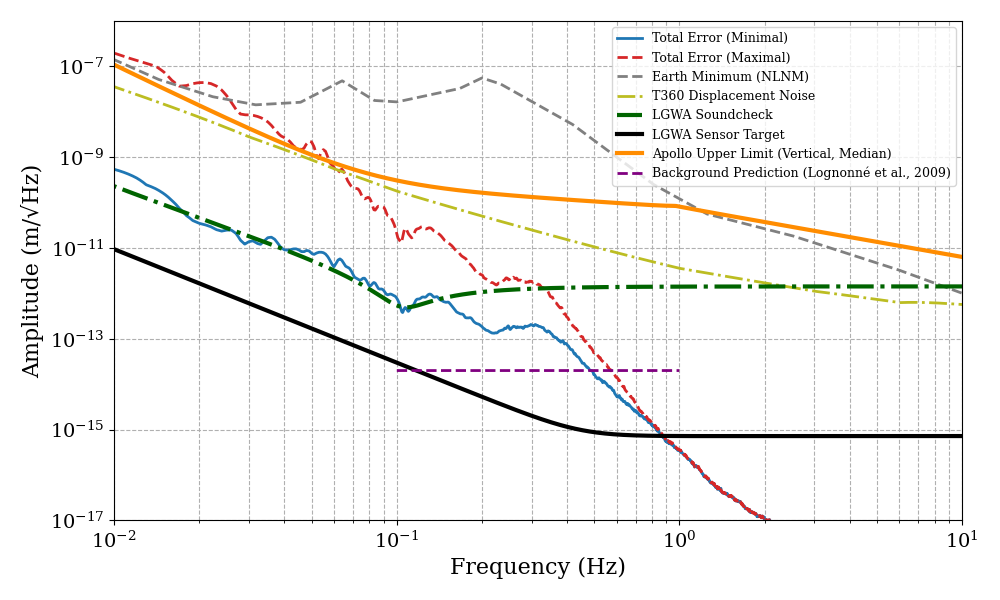} }}%
    \caption{(a) Individual contributions to the LGWA error signal. (b) Total error signal compared with reference noise levels for LGWA in both minimum and maximum-tilt scenarios.}%
    \label{fig:error_signal}
\end{figure}

As in the ET mode, each term in the summation of equation~(\ref{eq:contributions_error}) corresponds to a distinct noise contribution. Figure~\ref{fig:error_signal}a shows the individual contributions to the error signal, for both minimum and maximum tilt scenarios, while Figure~\ref{fig:error_signal}b presents the total error signal together with the target sensitivities of the LGWA sensor, LGWA Soundcheck, T360, and the Earth's minimum seismic background known as New Low-noise Model (NLNM).

In the maximum-tilt scenario, the total error signal \(X_{\mathrm{err}}\), is reduced to \(2 \times 10^{-11} \text{ m}/\sqrt{\text{Hz}}\) at \(100\,\mathrm{mHz}\), which will enable effective subtraction of the T360 sensor noise as it will be explained in the later section. This allows the intrinsic noise floor of the LGWA sensor \(N_{\mathrm{LGWA}}\), which reaches \(3 \times 10^{-14} \text{ m}/\sqrt{\text{Hz}}\) at \(100\,\mathrm{mHz}\), to be probed. At higher frequencies, the error signal decreases significantly, ensuring that control-induced noise does not limit the LGWA sensor testing performance. In the minimal tilt scenario, the total error signal lies below the LGWA sensor target above \(1\,\mathrm{Hz}\). At lower frequencies, it coincides with the LGWA Soundcheck sensitivity level. In the maximum-tilt case, the error signal remains below the LGWA sensor noise floor at higher frequencies as well, lies between the T360 and LGWA sensitivity between \(70\,\mathrm{Hz}\) and \(1\,\mathrm{Hz}\), and lies a bit above the T360 noise level at low frequencies.

Although the residual platform motion is not minimized in LGWA mode, it remains within acceptable bounds. A simple high-gain SISO controller for the \(X\) degree of freedom was adopted in LGWA mode, scaled by a factor compared to the ET mode $X$ DOF controller. This ensures a sufficiently high open-loop gain \(C \cdot P_{\rm ACT}\), enabling suppression of the error signal \(X_{\mathrm{err}}\) down to the sensor noise limits. The inclusion of the LGWA sensor, Soundcheck, and T360 sensitivity curves in Figure~\ref{fig:error_signal}b highlights the compatibility of GEMINI’s error signal suppression with the requirements for next-generation inertial sensor testing.

\subsubsection{Science Sensor Testing in LGWA Mode: Platform Motion, and Wiener Filtering}

Two types of sensors will be installed on the platform: the T360 seismometers, used for feedback control, and the LGWA sensors under test, which are more sensitive and have an intrinsic noise floor \(N_{\mathrm{LGWA}}\). The T360s will be mounted inside stage-1, providing the error signal for control, while the LGWA sensors will be mounted on the top of the platform, measuring out of loop, i.e., their signals are not used for platform control..

The LGWA sensor measures the residual platform motion given by equation~(\ref{eq:residual_LGWA}), along with tilt coupling, its own sensor noise, and electronics noise:

\begin{equation}
X_{\mathrm{LGWA}} = X_{\text{residual}} + \frac{g}{\omega^2} \theta
+ N_{\mathrm{LGWA}} + N_{\mathrm{ELECT}}^{(\mathrm{LGWA})}
\label{eq:lgwa_motion}
\end{equation}

For clarity, we simplify the expression for \(X_{\text{residual}}\) from equation~(\ref{eq:residual_LGWA}), focusing on the inertial sensing path without high-passing and neglecting actuator noise contributions consistent with the LGWA control objectives:

\begin{equation}
X_{\text{residual}} =
\frac{N_{\mathrm{seis}}}{1 + C \cdot P_{\rm ACT}}
- \frac{C \cdot P_{\rm ACT}}{1 + C \cdot P_{\rm ACT}}
\left( \frac{g}{\omega^2}\,\theta + N_{\mathrm{T360}} + N_{\mathrm{ELECT}}^{(\mathrm{T360})} \right)
\label{eq:residual_LGWA_abs}
\end{equation}

Substituting this expression into equation~(\ref{eq:lgwa_motion}), we obtain:
\begin{equation}
\begin{split}
X_{\mathrm{LGWA}} &=
\frac{N_{\mathrm{seis}}}{1 + C \cdot P_{\rm ACT}}
+ \frac{1}{1 + C \cdot P_{\rm ACT}} \frac{g}{\omega^2}\,\theta
- \frac{C \cdot P_{\rm ACT}}{1 + C \cdot P_{\rm ACT}}\, N_{\mathrm{T360}} \\
&\quad
- \frac{C \cdot P_{\rm ACT}}{1 + C \cdot P_{\rm ACT}}\, N_{\mathrm{ELECT}}^{(\mathrm{T360})}
+ N_{\mathrm{LGWA}} + N_{\mathrm{ELECT}}^{(\mathrm{LGWA})} \, .
\end{split}
\end{equation}

In the high control gain limit, we have:
\begin{equation}
\lim_{C \to \infty} X_{\mathrm{LGWA}}
= \bigl( N_{\mathrm{LGWA}} + N_{\mathrm{ELECT}}^{(\mathrm{LGWA})} \bigr)
 - \bigl( N_{\mathrm{T360}} + N_{\mathrm{ELECT}}^{(\mathrm{T360})} \bigr) \, .
\end{equation}

In this limit, the readout is the difference of the two sensor channels, so the readout–electronics terms $N_{\mathrm{ELECT}}^{(\mathrm{LGWA})}$ and $N_{\mathrm{ELECT}}^{(\mathrm{T360})}$ do not cancel. In practice, the in-loop T360 chain typically dominates via the control path. We will therefore (i) measure each readout chain’s electronics PSD and subtract its contribution in quadrature from the total measured power spectrum; and (ii) remove the coherent platform-motion contribution (details below). After these steps, the PSD of $N_{\mathrm{LGWA}}$ is estimated.

To overcome the limitation imposed by \(N_{\mathrm{T360}}\), Wiener filtering will be applied. Three LGWA horizontal seismometers will be placed on the platform table. One will be the target sensor, and the other two will be the witness sensors oriented at a right angle to each other. The Wiener filter produces a coherent estimate of the data of the target sensor using the data of the two witness sensors. We require two witness sensors to compensate for a potential misalignment, which might limit the coherence between a single witness sensor placed parallel to the target sensor.

The Wiener filter \(W(f)\) is constructed to minimize the mean-square error of the subtraction residual~\cite{Orf2007}. In the frequency domain, the residual obtained with two (or more) witness sensors can be written as
\begin{equation}
    X_{\rm res}(f) = X_{\mathrm{LGWA}}(f) - \vec W(f)\cdot \vec X_{\mathrm{wit}}(f)
\end{equation}
Ideally, the Wiener filter reduces the residual of the LGWA sensor data to
\begin{equation}
    X_{\rm res}(f) \approx N_{\mathrm{LGWA}}.
\end{equation}

This process effectively subtracts the control-induced T360 noise contribution, allowing direct observation of the LGWA sensor’s intrinsic noise floor.

As shown in Figure~\ref{fig:error_signal}, reaching the green curve already enables testing of Soundcheck sensors. However, achieving the LGWA sensor noise target (black curve) requires an additional 3--4 orders of magnitude improvement through effective Wiener filtering. Achieving this level of noise suppression critically depends on the level of coherence between sensors. For example, it is conceivable that even tiny deformations of the platform well below all resonance frequencies of its deformation modes might reduce coherence enough to limit the performance of the Wiener filter.

\vspace{1em}
\subsubsection{Summary and Implications for Sensor Testing in LGWA Mode}

The described measurement strategy enables GEMINI to test inertial sensors that are more sensitive than the T360. Key points include:
\begin{enumerate}
    \item High controller gain minimizes the error signal \(X_{\mathrm{err}}\).
    \item The LGWA sensor measures residual motion dominated by the difference between its own noise \(N_{\mathrm{LGWA}}\) and T360 noise \(N_{\mathrm{T360}}\).
    \item Wiener filtering removes the T360 contribution, revealing the LGWA sensor’s intrinsic performance.
    \item Optimal performance requires:
    \begin{itemize}
        \item Two witness and a target LGWA sensor to compensate for misalignments;
        \item A rigid platform;
        \item Sufficient dynamic range in the sensors;
        \item Enough data to calculate the correlations between channels with low enough statistical error.
    \end{itemize}
\end{enumerate}

\section{Conclusion}

GEMINI is conceived to advance seismic isolation and inter-platform control technologies for next-generation GW observatories. As the first underground research and development facility dedicated to GW technologies, GEMINI addresses the stringent performance requirements of both the Einstein Telescope and the Lunar Gravitational-Wave Antenna through its dual-mode control architecture and cryogenic testing capabilities.

In what we call \emph{ET mode}, the system prioritizes minimizing residual platform motion to provide stable suspension points for interferometric optics. A key feature enabling this is the Suspension Platform Interferometer, which enforces common motion across platforms to form an optically rigid body. This rigid reference is essential for suppressing excess alignment and length noise in the auxiliary interferometric degrees of freedom such as the recycling cavities. Our analysis demonstrates that the SPI control loop substantially reduces differential motion, confirming its utility in meeting the most challenging stability goals.

In the \emph{LGWA mode}, the control objective shifts from minimizing platform motion to minimizing the error signal. The most important difference here is that tilt-to-horizontal coupling does not interfere with how much the horizontal error signal can be suppressed. This approach establishes an ultra-quiet inertial reference frame suitable for evaluating cryogenic seismometers with performance exceeding that of today's state-of-the-art. The GEMINI environment allows testing under conditions closely resembling those on the Moon, enabling realistic huddle tests of LGWA payloads prior to deployment. At high feedback gain, the science sensor (e.g., LGWA prototype) effectively measures the difference between its own intrinsic noise and the noise of the control sensor (T360 one). The latter appears correlated between to-be-tested sensors placed on the table and can be subtracted via Wiener filtering.

GEMINI’s design targets picometer-level platform displacements. Ongoing upgrades and future optimization will focus on further reducing residual motion, improving coherence in inter-platform control, and validating the system’s performance against ET and LGWA operational requirements. This dual-mode architecture, supported by active and passive control techniques and validated in a realistic underground and cryogenic environment, makes GEMINI a critical testbed for technologies that will underpin the next generation of GW observatories.

\paragraph{Future Work}-- Future developments will focus on several key areas:
\begin{itemize}
    \item Determining the required SPI sensitivity is nontrivial — it requires modeling of how residual platform motion couples into the control of auxiliary DOFs at the ET vertices (e.g., PRC, SEC) taking experience with current GW detectors into account.
    \item Time-Domain Simulations and Validation: Extending the current frequency-domain analysis to time-domain simulations with Lightsaber \cite{AndHar2021}, providing a more comprehensive validation of controller performance under realistic operating conditions and transient disturbances.
    \item Controller Optimization: Exploring alternative control strategies, including the design of machine learning (ML)-based controllers, to optimize performance in both ET and LGWA modes. ML approaches may offer adaptive control schemes that improve stability and disturbance rejection in complex environments. Here, time-domain simulations will play a key role (chapter 4 of~\cite{And2023, Buchlietal2025}).
    \item Environmental Noise Coupling Analysis: Developing and validating comprehensive models of environmental noise coupling to platform motion, including seismic, acoustic, and thermal effects. This will inform sensor placement, shielding strategies, and feedback and feedforward controller design to further suppress environmental disturbances.
\end{itemize}

GEMINI is posed to advance technologies required by the Einstein Telescope and for characterizing the cryogenic inertial sensors essential to the Lunar Gravitational-Wave Antenna. Its flexible control architecture, coupled with advanced sensing and feedback systems, positions GEMINI as a key technology demonstrator for future GW detection on Earth and beyond.

\section*{Acknowledgments}
We thank Jeff Kissel (LIGO Hanford Observatory) for insightful discussions on HAM-ISI; several aspects of the GEMINI control strategy benefited directly from his guidance. We are also grateful to Carlo Bucci (INFN - Laboratori Nazionali del Gran Sasso (LNGS)) for helpful conversations on infrastructure. 

This research was supported in part by the Deutsche Forschungsgemeinschaft (DFG, German Research Foundation) under Germany's Excellence Strategy---EXC 2121 ``Quantum Universe''---390833306.
This project has received funding from the European Research Council (ERC) under the European Union's Horizon 2020 research and innovation programme (grant agreement No. 865816)

\printbibliography

\end{document}